\documentclass[12pt,a4paper]{article}

\usepackage[margin=1in]{geometry}
\usepackage[sort&compress,square,numbers]{natbib}
\usepackage{lipsum,times,graphicx}
\usepackage[labelfont=bf]{caption}
\usepackage{authblk}
\usepackage{threeparttable}
\usepackage{lineno}
\usepackage{amsmath,amssymb,amsfonts}
\usepackage{amsthm}
\usepackage{mathrsfs}
\usepackage{algorithm}
\usepackage{algorithmicx}
\usepackage[version=4]{mhchem}
\usepackage{multirow}
\usepackage{booktabs}
\usepackage{soul}
\usepackage{color, xcolor}
\usepackage[normalem]{ulem}
\usepackage{xurl}
\usepackage[colorlinks,
		linkcolor=blue,
		anchorcolor=blue,
		citecolor=blue,
		urlcolor=blue]{hyperref}

\usepackage{multibib}
\newcites{SI}{Supplementary References}

\usepackage{titletoc}

\date{} 

\makeatletter
\newcommand{\startappendix}{%
  \appendix
  \small

  \renewcommand\section{\@startsection{section}{1}{\z@}%
    {-3.5ex \@plus -1ex \@minus -.2ex}%
    {2.3ex \@plus.2ex}%
    {\normalfont\large\bfseries}}
    
  \renewcommand\subsection{\@startsection{subsection}{2}{\z@}%
    {-3.25ex\@plus -1ex \@minus -.2ex}%
    {1.5ex \@plus .2ex}%
    {\normalfont\normalsize\bfseries}}
    
  \renewcommand\subsubsection{\@startsection{subsubsection}{3}{\z@}%
    {-3.25ex\@plus -1ex \@minus -.2ex}%
    {1.5ex \@plus .2ex}%
    {\normalfont\normalsize\bfseries}}
  
  \captionsetup{font=small}
  \AtBeginEnvironment{tabular}{\small}%
  \AtBeginEnvironment{table}{\small}%
}
\makeatother

\title{
\fontsize{19}{22}\selectfont\bfseries Accelerating inverse materials design using generative diffusion models with reinforcement learning
}

\author[1,2*]{Junwu Chen}
\author[1,2]{Jeff Guo}
\author[1,2]{Edvin Fako}
\author[1,2*]{Philippe Schwaller}
\affil[1]{Laboratory of Artificial Chemical Intelligence (LIAC), 
  Institute of Chemical Sciences and Engineering, 
  Ecole Polytechnique F\'{e}d\'{e}rale de Lausanne (EPFL), 
  Lausanne, Switzerland}
\affil[2]{National Centre of Competence in Research (NCCR) Catalysis, 
  Ecole Polytechnique F\'{e}d\'{e}rale de Lausanne (EPFL), 
  Lausanne, Switzerland}
\affil[*]{Corresponding authors. Email: junwu.chen@epfl.ch, philippe.schwaller@epfl.ch}

\begin{document}
\maketitle
\clearpage

\section*{ABSTRACT}

Diffusion models promise to accelerate material design by directly generating novel structures with desired properties, but existing approaches typically require expensive and substantial labeled data ($>$10,000) and lack adaptability. Here we present MatInvent, a general and efficient reinforcement learning workflow that optimizes diffusion models for goal-directed crystal generation. For single-objective designs, MatInvent rapidly converges to target values within 60 iterations ($\sim$ 1,000 property evaluations) across electronic, magnetic, mechanical, thermal, and physicochemical properties. Furthermore, MatInvent achieves robust optimization in design tasks with multiple conflicting properties, successfully proposing low-supply-chain-risk magnets and high-$\kappa$ dielectrics. Compared to state-of-the-art methods, MatInvent exhibits superior generation performance under specified property constraints while dramatically reducing the demand for property computation by up to 378-fold. Compatible with diverse diffusion model architectures and property constraints, MatInvent could offer broad applicability in materials discovery.

\textbf{Keywords:} reinforcement learning, diffusion generative model, inverse materials design
\clearpage

\section{Introduction} \label{introduction}

The development of novel functional materials is pivotal for accelerating scientific progress in various fields such as catalysis, microelectronics, and renewable energy \cite{mou2023bridging, yao2023machine, molesky2018inverse}. The key objective is to identify property-optimal candidates within an enormous design space. Existing methods include iterative experimental trial-and-error \cite{raccuglia2016machine, inverse-design-materials-review} and high-throughput screening \cite{szymanski2023autonomous, tom2024self}. However, the brute-force screening of all possible materials is prohibitive, and is a challenge circumvented through expert-defined search spaces. Although this approach has offered success in discovering novel materials, manually constraining the search space could introduce negative bias. 

Recently, generative models \cite{wang2025leveraging, cdvae, xiao2023invertible, miller2024flowmm, gruver2024fine}, particularly diffusion models \cite{diffcsp, mattergen, cao2024space, yang2024scalable, levysymmcd}, have emerged as promising frameworks for generating novel and theoretically stable inorganic crystal structures spanning the entire periodic table. Several methods, such as conditional generation by classifier-free guidance \cite{mattergen}, have been proposed to steer diffusion models toward generating materials with targeted properties \cite{guo2025ab, okabe2025structural, mal2025diffcrysgen}. Nevertheless, these methods require substantial pre-existing labeled data for model fine-tuning, limiting their generalizability and flexibility across diverse inverse design tasks.

Reinforcement learning (RL) provides a framework for optimizing generative models by iterative exploration of complex problem spaces based on feedback rewards, with potential to improve generation quality and controllability \cite{deepseek-r1, dpok, ddpo}. This approach decouples learning from dense annotations by leveraging sparse or indirect reward signals, requiring substantially fewer labeled data compared to supervised fine-tuning \cite{dpok, ddpo}. Notably, RL has become a mainstream strategy for optimizing SMILES-based language models to accomplish goal-directed molecular generation \cite{reinvent, neil-segler-brown, popova-rnn, augmented-memory, saturn}. Although several RL approaches have been proposed for crystal structure prediction \cite{zamaraeva2023reinforcement, govindarajan2024learning} and composition generation of metal oxides \cite{karpovich2024deep}, RL frameworks for optimizing diffusion models in inorganic materials design remain scarce \cite{matinvent, crystalgym, crystalformer-rl}.

This work proposes MatInvent, a versatile and efficient RL workflow for optimizing pre-trained diffusion models toward objective-driven crystal generation. By framing denoising generation as a multi-step decision-making problem, MatInvent leverages policy optimization with reward-weighted Kullback–Leibler (KL) regularization, including experience replay and diversity filters to enhance sample efficiency and diversity. For single-objective optimization, MatInvent demonstrates remarkable performance and flexibility across various material design tasks encompassing electronic, magnetic, mechanical, physicochemical, thermal, and synthesizability properties. Compared to conditional generation of MatterGen, our RL approach substantially reduces the requirement for labeled data while exhibiting enhanced generative performance under target property constraints. Furthermore, MatInvent achieves robust optimization in design tasks with multiple competing objectives, successfully designing magnets with low supply-chain risk and high-$\kappa$ gate dielectrics. This versatility makes our approach highly appealing to researchers in materials science, chemistry, and catalysis.

\begin{figure}[htp]
\centering
\includegraphics[width=1.0\textwidth]{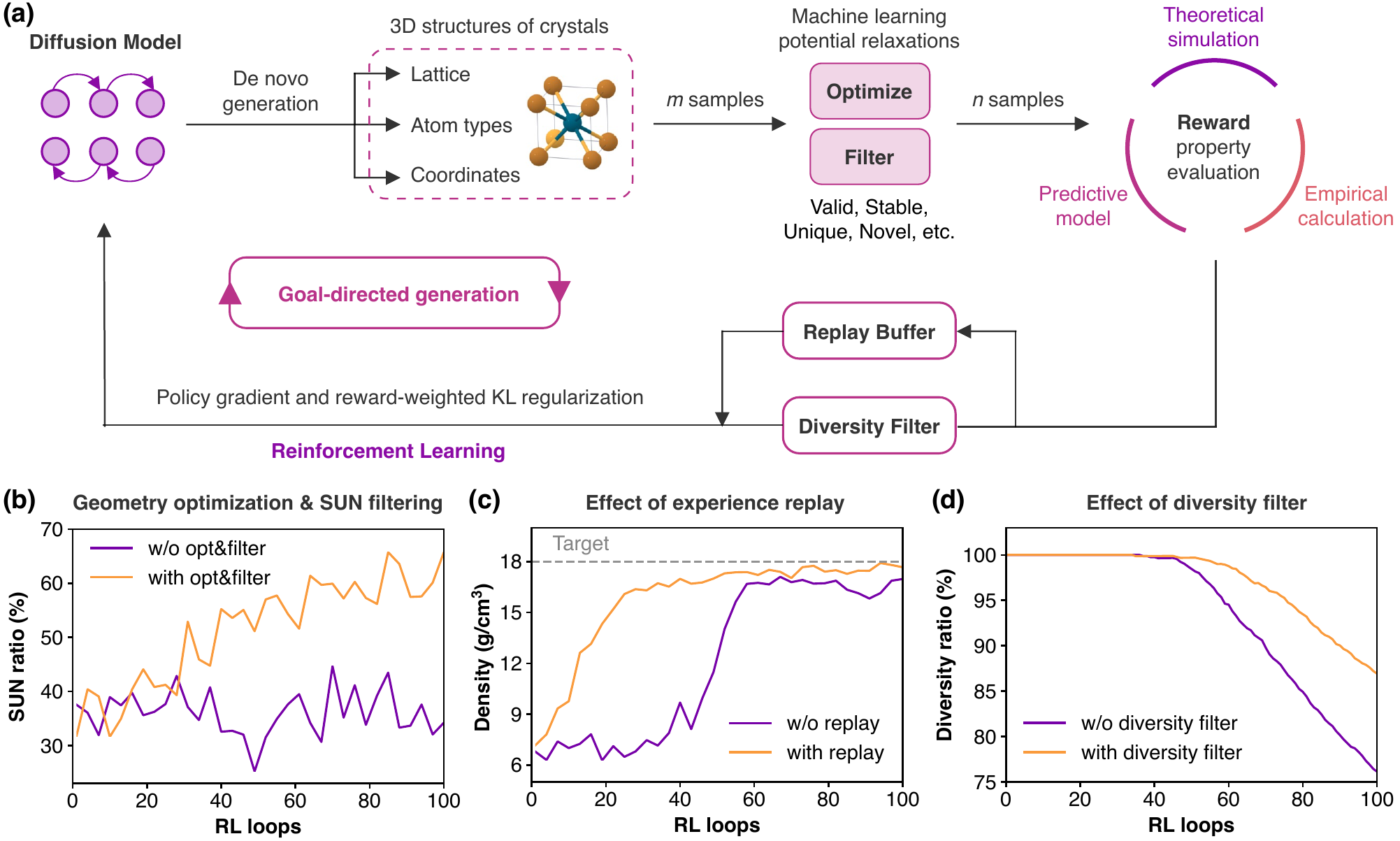}
\caption{\textbf{MatInvent workflow for goal-directed material generation.} 
(a) The schematic overview of MatInvent methodology. In each reinforcement learning (RL) iteration, the diffusion model acts as the RL agent to generate a batch of 3D crystal structures, which are subsequently geometrically optimized using machine learning potentials. Only valid, Stable, Unique, and Novel (SUN) structures are retained after filtering, proceeding to target property evaluation and reward assignment. High-reward samples are then used to fine-tune the diffusion model by policy optimization with reward-weighted Kullback–Leibler (KL) regularization, aided by experience replay and diversity filter to enhance sample efficiency and diversity.
(b) The impact of geometry optimization (opt) and SUN filtering before property evaluation on the SUN ratio of generated structures during the RL process targeting a density of 18.0 g/cm$^3$.
(c) The effect of experience replay on the optimization efficiency of RL process targeting a density of 18.0 g/cm$^3$.
(d) The role of diversity filter in the composition diversity of generated structures during the RL process with a target density of 18.0 g/cm$^3$.
}
\label{fig:rl_pipeline}
\end{figure}

\section{Results}
\subsection{Reinforcement learning pipeline}
MatInvent is an RL workflow designed for goal-directed generation of crystalline materials (Fig. \ref{fig:rl_pipeline}). In the pipeline, the diffusion model acts as the RL agent that generates novel 3D crystal structures through a $T$-step reverse denoising process on atomic types, atomic coordinates, and lattice matrices \cite{diffcsp, mattergen}. The denoising process of the diffusion model can be reframed as a $T$-step Markov decision process (MDP) \cite{dpok, ddpo} for our online RL algorithms (see Methods). We denote the diffusion model before RL fine-tuning as the prior, which was pre-trained on large-scale unlabeled datasets of crystal structures (e.g., Alex-MP \cite{mattergen}) and can generate diverse crystalline materials spanning over 80 elements. In each RL iteration, the diffusion model randomly generates a batch of $m$ crystal structures. The generated structures undergo geometry optimization using universal ML interatomic potentials (MLIP) \cite{mattersim} and their energy above hull ($E_{hull}$) is calculated. Only crystal structures that are thermodynamically Stable ($E_{hull} <$ 0.1 eV/atom), Unique, and Novel (SUN) \cite{mattergen} are retained after filtering, in which $n$ samples are randomly selected for property evaluation and assigned corresponding rewards. The material properties and rewards can be obtained through theoretical simulations, ML predictions, and empirical calculations. The top $k$ samples ranked by reward are used to fine-tune the diffusion model based on policy optimization with reward-weighted KL regularization (see Methods).
The KL regularizer between the pre-trained and fine-tuned models is incorporated into the RL objective function to prevent reward overfitting while preserving the material knowledge acquired during pre-training \cite{dpok}.
Moreover, experience replay and the diversity filter are respectively employed to improve optimization efficiency and sample diversity of RL process, enabling faster convergence toward the target while generating novel and diverse crystal structures. Experience replay is used to improve the stability and efficiency of learning by storing past high-reward crystals in a replay buffer and reusing them during RL fine-tuning \cite{experience-replay, augmented-memory}. The diversity filter imposes a linear penalty on the reward of non-unique crystal structures based on the number of previous occurrences \cite{diversity-filter, thomas2022augmented}. Specifically, crystals with the same structure or composition as previously generated samples are assigned reduced rewards and subsequently removed from the replay buffer, thereby encouraging the diffusion model to explore unseen material space. Notably, MatInvent is a general-purpose RL workflow that is compatible with different diffusion model architectures (Supplementary Information section \ref{si:sec:general}). Unless otherwise specified, all experiments in this work use the MatterGen \cite{mattergen} framework as the diffusion model in the RL process.

To investigate the importance of individual components in our RL workflow, ablation studies were conducted using the material design task with a target density of 18.0 g/cm$^3$ (Supplementary Information section \ref{si:sec:ablation}). Two metrics, the SUN ratio and composition diversity ratio, are defined to evaluate the generation quality and diversity of material structures from diffusion models (Supplementary Information section \ref{si:sec:metrics}). As shown in Fig. \ref{fig:rl_pipeline}b and Supplementary Information section \ref{si:sec:ablation:optf}, MLIP-based geometry optimization and SUN filtering prior to property evaluation improve both the SUN ratio and composition diversity of the generated structures during the RL process. As depicted in Fig. \ref{fig:rl_pipeline}c and Supplementary Information section \ref{si:sec:ablation:replay}, experience replay enhances the RL optimization efficiency, enabling faster convergence to the target property value with fewer property evaluations. This is particularly important for material properties that have expensive evaluation costs. Moreover, the diversity filter can encourage diffusion models to explore different chemical systems and achieve a higher diversity ratio of chemical compositions during the RL process (Fig. \ref{fig:rl_pipeline}d and Supplementary Information section \ref{si:sec:ablation:df}). This facilitates the design of diverse crystal structures with target properties and prevents the RL optimization from stagnating in local minima.

\subsection{Single property optimization}

\begin{figure}[htp]
\centering
\includegraphics[width=1.0\textwidth]{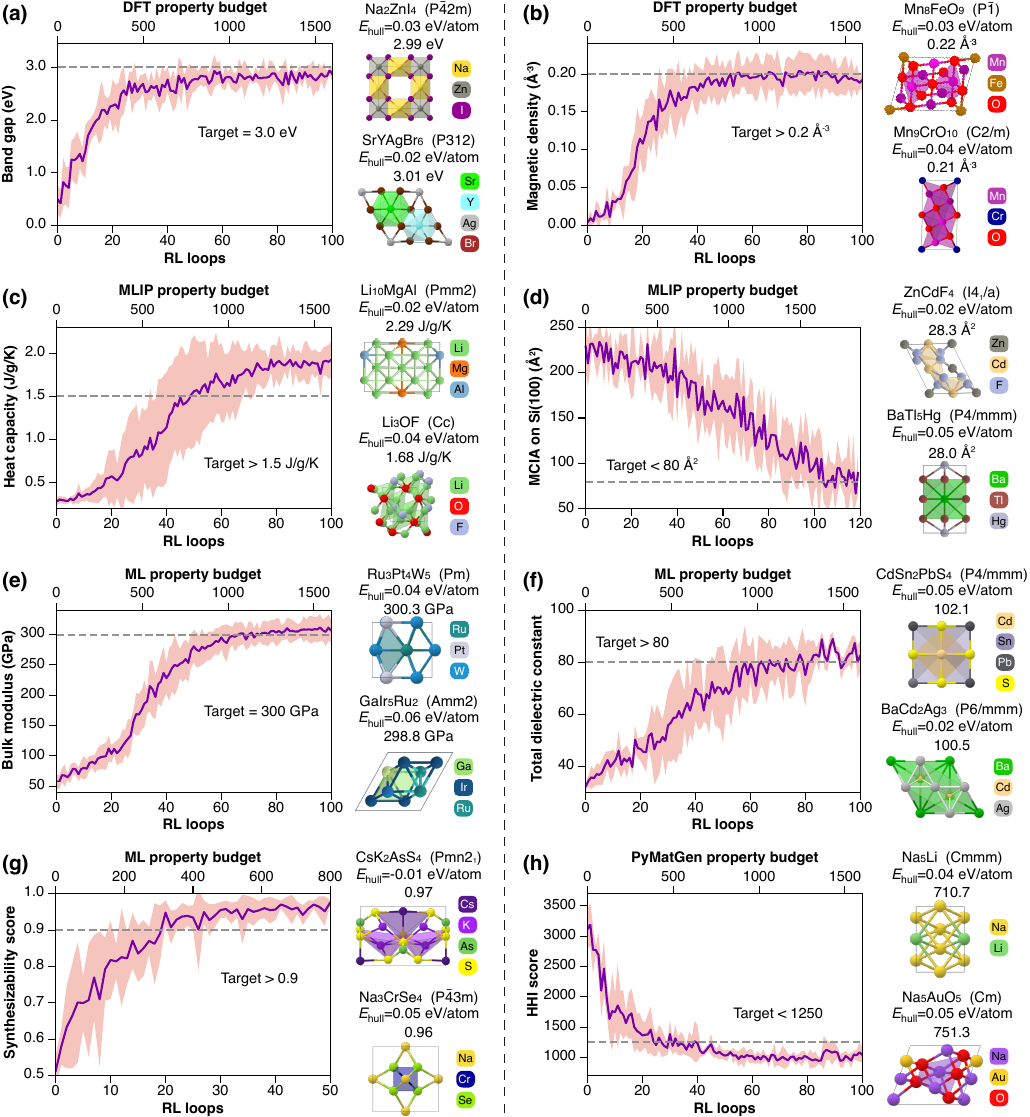}
\caption{\textbf{MatInvent performance on single property optimization.}
The optimization curves (left) for reinforcement learning (RL) and visualizations of some generated crystal structures (right) on different inverse design tasks with a single target property:
(a) band gap equal to 3.0 eV;
(b) magnetic density higher than 0.2 $\text{\AA}^{-3}$;
(c) specific heat capacity exceeding 1.5 J/g/K;
(d) minimal co-incident area (MCIA) below 80 $\text{\AA}^{2}$ on the Si(100) substrate;
(e) bulk modulus of 300 GPa;
(f) total dielectric constants exceeding 80;
(g) synthesizability score higher than 0.9;
and (h) Herfindahl–Hirschman index (HHI) score below 1250.
Ten repeat experiments were performed for tasks c–h, while three for tasks a and b. The curves show the mean of repeated experiments while the shading represents standard deviation.
}
\label{fig:spo}
\end{figure}

In numerous applications, such as energy storage, superconductivity, and electronic devices, the primary demand lies in designing novel materials with targeted or enhanced properties. MatInvent was evaluated on different inverse design tasks for single property optimization. These tasks encompass various properties of inorganic materials, including electronic, magnetic, mechanical, thermal, physicochemical, and synthesizability characteristics. The property values and corresponding rewards are derived from density functional theory (DFT) calculations (Fig. \ref{fig:spo} a and b), MLIP simulations (Fig. \ref{fig:spo} c and d), or ML prediction models (Fig. \ref{fig:spo} e, f, and g). The first task (Fig. \ref{fig:spo}a) aims to generate materials with a target band gap of 3.0 eV, a key property for light-emitting devices \cite{smith2010semiconductor}, photocatalysis \cite{xu2012enhancing}, and wide-bandgap semiconductor \cite{zhang2025wide}. In the second task (Fig. \ref{fig:spo}b), the goal is to design materials with magnetic densities higher than 0.2 $\text{\AA}^{-3}$, a prerequisite for permanent magnets \cite{mattergen}. The third task (Fig. \ref{fig:spo}c) involves generating novel inorganic compounds with specific heat capacities exceeding 1.5 J/g/K, which is crucial for thermal energy storage and high-temperature protection materials \cite{heat_capacity}. The fourth task (Fig. \ref{fig:spo}d) focuses on designing novel crystal structures with strong epitaxial matching to the commercially dominant Si(100) substrate, requiring a minimal co-incident area (MCIA) below 80 $\text{\AA}^{2}$ \cite{mcia1}. A lower MCIA indicates a higher degree of matching between the thin-film material and the substrate, which is crucial for material synthesis techniques such as chemical vapor deposition and sputtering \cite{mcia1}. The fifth task (Fig. \ref{fig:spo}e) targets the generation of materials with a high bulk modulus of 300 GPa, an essential property for superhard and aerospace materials \cite{superhard}. The sixth task (Fig. \ref{fig:spo}f) focuses on discovering new materials with high total dielectric constants exceeding 80, important for applications such as electronic devices and supercapacitors \cite{dielectrics1, dielectrics2}. The seventh task (Fig. \ref{fig:spo}g) investigates the generation of materials with high synthesizability scores based on feedback from the ML model \cite{syn_score}, aiming to design novel and experimentally synthesizable materials. Finally, the eighth task (Fig. \ref{fig:spo}h) is to design new materials with low supply chain risk, requiring a Herfindahl–Hirschman index (HHI) score \cite{hhi} below 1250 directly computed through PyMatGen \cite{pymatgen}. Further details on the RL experiments and reward calculation are provided in Supplementary Information sections \ref{si:sec:spo_exp} and \ref{si:sec:spo_reward}, while the methods for material property evaluation are described in Supplementary Information section \ref{si:sec:property}.

As shown in Fig. \ref{fig:spo}a–h, across all tasks, the average property values of the generated materials progressively approach the target values with successive RL iterations. Remarkably, within 60 iterations and $\sim$1000 property evaluation calls, the average property values converge to their targets for most tasks. Six more single-objective design tasks were also explored (Supplementary Information section \ref{si:sec:spo_more}), involving shear modulus, Young's modulus, Pugh ratio, formation energy, crustal abundance, and price. Moreover, the property distributions of the SUN structures generated by the RL fine-tuned model displayed a clear shift, and became more concentrated around the target values compared with the pre-trained model (Supplementary Information section \ref{si:sec:spo_dis}). This confirms that MatInvent can optimize diffusion models and steer their generative distribution toward regions of materials with desired properties. As depicted in Supplementary Information section \ref{si:sec:spo_sun}, most RL fine-tuned models exhibited higher SUN ratios ($>$ 45 \%) relative to the initial pretrained model (38.7 \%), which can be attributed to MLIP-based structure optimization and SUN filtering prior to property evaluation. All results demonstrate that MatInvent is an efficient and general RL framework for diffusion models in single-property inverse design tasks.

\begin{figure}[htp]
\centering
\includegraphics[width=1.0\textwidth]{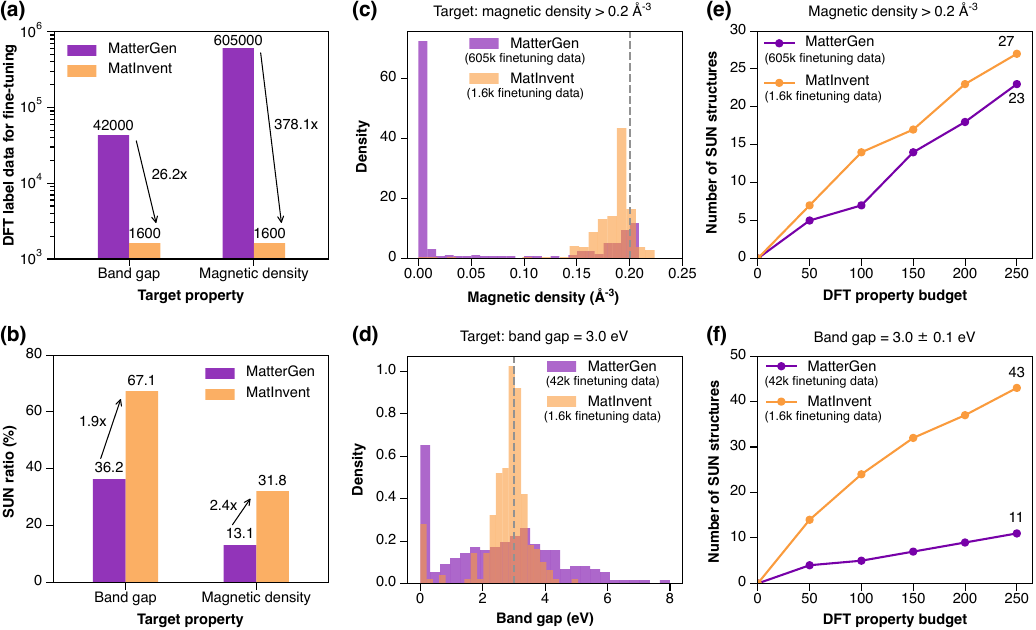}
\caption{\textbf{Comparison between conditional generation and reinforcement learning.}
(a) Number of DFT-labeled data used for model fine-tuning in the MatInvent workflow and conditional generation of MatterGen across two inverse design tasks.
(b) SUN ratios of generated structures from MatterGen conditional generation and RL-finetuned diffusion model following the MatInvent workflow.
Probability density distributions of property values of SUN structures generated by RL-finetuned diffusion models and MatterGen's conditional generation, respectively, for inverse design targets of (c) magnetic density higher than 0.2 $\text{\AA}^{-3}$ and (d) band gap of 3.0 eV.
Number of SUN structures satisfying property requirements discovered by MatterGen conditional generation and RL-finetuned diffusion models within 250 DFT property calculations, for targets of (e) magnetic density higher than 0.2 $\text{\AA}^{-3}$ and (f) band gap of $3 \pm 0.1$ eV.
}
\label{fig:mattergen}
\end{figure}

We further compared MatInvent with the state-of-the-art conditional generation method (MatterGen \cite{mattergen}), on two specific tasks: targeting materials with bandgaps of 3.0 eV and magnetic densities exceeding 0.2 $\text{\AA}^{-3}$. For a fair comparison, all RL experiments used the same unconditional MatterGen model pre-trained on Alex-MP-20 dataset as the initial model \cite{mattergen}, and more details are in Supplementary Information section \ref{si:sec:spo_mattergen}. For MatterGen's conditional generation, the pre-trained model with adapter modules undergoes fine-tuning on pre-existing and DFT labeled datasets, subsequently applying classifier-free guidance to steer crystal generation toward the desired objectives \cite{mattergen}. As illustrated in Fig. \ref{fig:mattergen}a, MatterGen’s conditional generation method requires 42,000 and 605,000 DFT-labeled data points for fine-tuning on the two tasks \cite{mattergen}, respectively, whereas MatInvent needs only 1,600 DFT calculations to obtain rewards for 100 RL iterations. MatInvent substantially reduces the expensive DFT computational costs required for model fine-tuning by factors of 26 and 378 on the two tasks, respectively (Fig. \ref{fig:mattergen}a). Moreover, the RL-finetuned model demonstrates approximately twice the SUN ratio compared to the conditional generation of MatterGen (Fig. \ref{fig:mattergen}b). In Fig. \ref{fig:mattergen}c and d, the property distributions of SUN structures generated by the RL-finetuned model are more concentrated around target values in both tasks, compared to those from MatterGen’s conditional generation. We also evaluated the performance of MatInvent against MatterGen's conditional generation in discovering SUN structures that satisfy stringent property requirements under limited DFT calculation budgets. As shown in Fig. \ref{fig:mattergen}e, the RL-finetuned model identified 27 SUN structures with magnetic densities exceeding 0.2 $\text{\AA}^{-3}$ within a budget of 250 DFT property calculations, outperforming MatterGen conditional generation (23 structures). Figure \ref{fig:mattergen}f reveals that the RL-finetuned model discovered 43 SUN structures with band gaps of 3.0 $\pm$ 0.1 eV after 250 DFT property calculations, substantially surpassing the conditional generation of MatterGen (11 structures). It is worth noting that tasks with narrow property range constraints present comparable challenges to those with extreme target properties. All results demonstrate that MatInvent achieves improved goal-directed crystal generation performance, while significantly reducing DFT computational burden.
This arises because RL directly optimizes the diffusion model to maximize rewards, concentrating generation in high-reward regions, whereas conditional generation requires learning the complete conditional probability distribution over the target property.

\subsection{Multiple property optimization}

\begin{figure}[htp]
\centering
\includegraphics[width=0.9\textwidth]{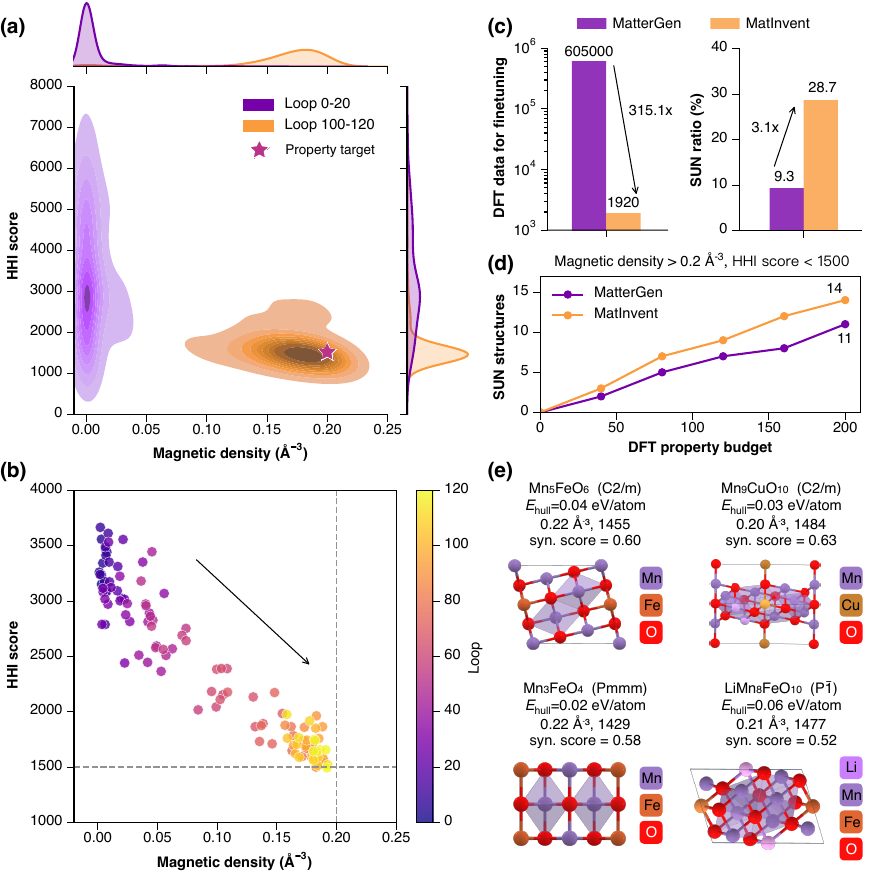}
\caption{\textbf{Designing permanent magnets with low supply chain risk.}
(a) Property distribution of SUN structures generated during the initial (0–20 loops) and final (100–120 loops) stages of RL process.
(b) Mean values of target properties of SUN structures generated in each RL iteration.
(c) Amount of DFT-labeled data used for model fine-tuning (left) and SUN ratios of generated structures (right) for MatterGen conditional generation and MatInvent workflow.
(d) Number of SUN structures satisfying property requirements found by MatterGen conditional generation and RL-finetuned diffusion models within 200 DFT property calculations, for targets with magnetic density above 0.2 $\text{\AA}^{-3}$ and HHI score below 1500.
(e) Visualizations of some SUN structures generated by RL-finetuned diffusion models, along with their chemical formula, space group, energy above hull ($E_{hull}$), magnetic density, HHI score, and synthesizability score.
}
\label{fig:mag_vs_hhi}
\end{figure}

Most material design problems require finding structures that satisfy multiple property constraints. Two tasks were designed to evaluate the performance of MatInvent in the simultaneous optimization of multiple material properties. The first task focuses on designing novel permanent magnets with low supply chain risk, aiming to avoid the utilization of rare-earth elements \cite{mattergen}. This task can be formulated as satisfying two property constraints: (1) magnetic density higher than 0.2 $\text{\AA}^{-3}$, and (2) Herfindahl–Hirschman index (HHI) score below 1500. An HHI score below 1500 is considered indicative of low supply chain risk \cite{hhi}. In the RL experiments, the DFT method was employed to determine the magnetic densities of the generated structures, and PyMatGen \cite{pymatgen} was utilized to compute their HHI scores. The minimum between the scaled values of magnetic density and HHI score served as the reward for each sample during the online RL process, thereby facilitating the simultaneous optimization of both target properties (Supplementary Information section  \ref{si:sec:mpo_reward}). As illustrated in Fig. \ref{fig:mag_vs_hhi}b, the average values of both properties for the generated SUN structures gradually approached the target region with successive RL iterations, ultimately converging near the desired values after 100 iterations. In contrast to the initial phase of RL (loops 0-20), the distribution of both properties for SUN structures generated during loops 100-120 exhibited a pronounced shift and became narrowly concentrated around the target values, as demonstrated in Fig. \ref{fig:mag_vs_hhi}a. These findings demonstrate that MatInvent can iteratively optimize the diffusion model and its generation distribution for two competing material properties.

As depicted in Fig. \ref{fig:mag_vs_hhi}c, MatInvent required only 1,920 DFT property calculations for 120 RL iterations, representing a 315-fold reduction compared to the 605,000 DFT-labeled data points used for fine-tuning in MatterGen's conditional generation \cite{mattergen}. Moreover, we compared the performance between the RL-finetuned diffusion model and MatterGen's conditional generation in discovering SUN structures that meet two property requirements under limited DFT computation budgets (Supplementary Information section \ref{si:sec:mpo_mattergen}). As shown in Fig. \ref{fig:mag_vs_hhi}d, the RL-finetuned diffusion model identified 14 SUN structures satisfying both property requirements under 200 DFT property calculations, outperforming MatterGen's conditional generation (11 SUN structures). Of the 14 SUN structures found by MatInvent, 78.6 \% (n=11) exhibit ML-predicted synthesizability scores above 0.5, indicating potential experimental feasibility \cite{syn_score}. Some of these structures are presented in Fig. \ref{fig:mag_vs_hhi}e. Overall, all results establish that MatInvent is highly efficient for inverse design tasks with multiple objectives, achieving superior crystal generation performance, while drastically reducing DFT computational costs.

\begin{figure}[htp]
\centering
\includegraphics[width=1.0\textwidth]{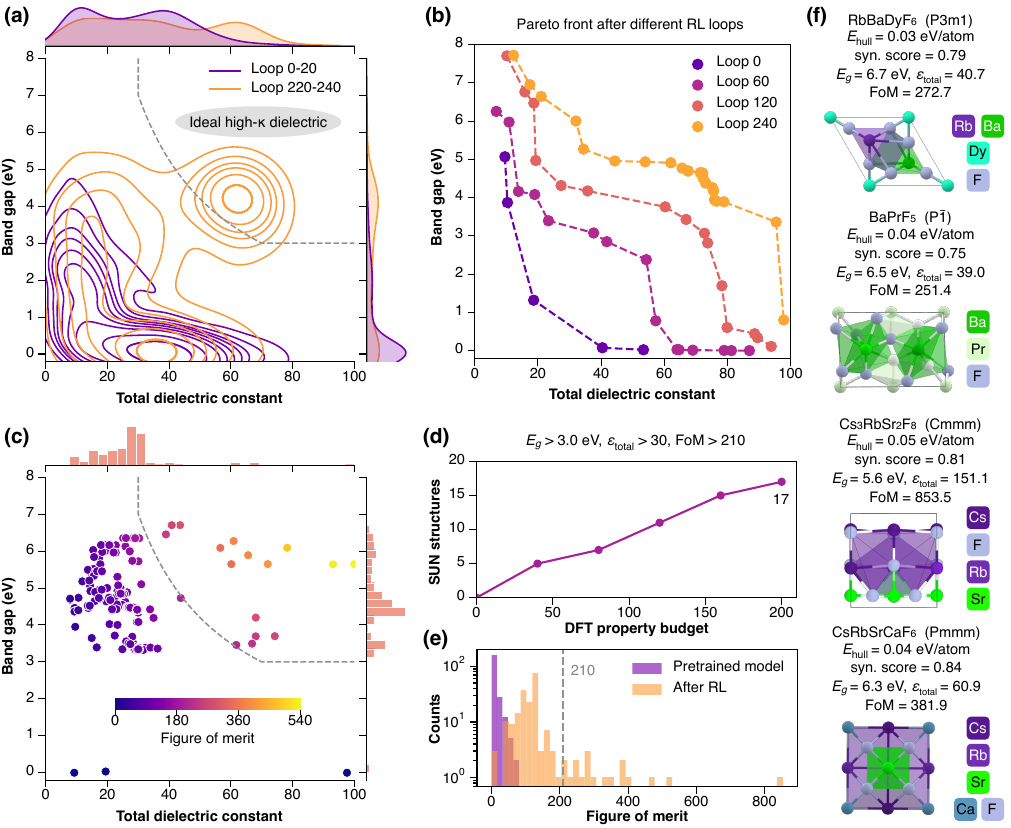}
\caption{\textbf{Designing novel high-$\kappa$ dielectrics.}
(a) Property distribution of SUN structures generated during the initial (0–20 loops) and final (220–240 loops) stages of RL process.
(b) Evolution of Pareto fronts across RL iterations for two conflicting material properties: dielectric constant and band gap.
(c) DFT-calculated property distribution of SUN structures generated by the RL-finetuned diffusion model, which were ranked and selected based on ML predictions.
(d) Number of SUN structures satisfying property requirements found by RL-finetuned diffusion models within 200 DFT property calculations, for objectives of band gap ($E_g$) exceeding 3.0 eV, total dielectric constant ($\varepsilon_{\text {total }}$) surpassing 30, and figure of merit (FoM) higher than 210.
(e) Distribution of DFT-computed figure of merit for generated structures by the pre-trained and RL-finetuned diffusion models.
(f) Visualizations of some SUN structures generated by RL-finetuned diffusion models, along with their chemical formula, space group, energy above hull ($E_{hull}$), synthesizability score, $E_g$, $\varepsilon_{\text {total }}$, and FoM.
}
\label{fig:mpo_dielectrics}
\end{figure}

The second task aims to design novel high-$\kappa$ dielectrics, critical components in numerous microelectronic devices, including central processing units (CPU), dynamic random-access memory (DRAM), and high-frequency antennas \cite{dielectrics1, dielectrics2}. Their performance depends on an intricate balance between a high dielectric constant and a wide bandgap, two inversely correlated characteristics that rarely co-exist within a single material \cite{dielectrics1, dielectrics2}. Moreover, a high figure of merit (FoM) is desirable to suppress tunneling current, while one top experimentally reported high-$\kappa$ dielectric t-HfO$_2$ exhibits an FoM of approximately 210 \cite{dielectrics2}. Consequently, the second task can be formulated with three optimization objectives: band gap ($E_g$) exceeding 3.0 eV, total dielectric constant ($\varepsilon_{\text {total }}$) surpassing 30, and FoM higher than 210 \cite{dielectrics2}. Given the computational expense of DFT property evaluation, ML models were employed to predict $E_g$, $\varepsilon_{\text {total }}$, and corresponding FoM ($=E_g \times \varepsilon_{\text{total}}$) of the crystal structures generated during the RL process (Supplementary Information section \ref{si:sec:mpo_reward}). As illustrated in Fig. \ref{fig:mpo_dielectrics}a, the structures generated during early-stage RL (0-20 iterations) fall into two primary categories: wide $E_g$ but low $\varepsilon_{\text {total }}$, and narrow $E_g$ but high $\varepsilon_{\text {total }}$. After 220 RL iterations, the property distributions of generated SUN structures exhibited a pronounced shift toward the target region with high $E_g$ and $\varepsilon_{\text {total }}$, compared to the initial phase of RL (0-20 iterations). Correspondingly, the Pareto frontier was progressively optimized during the RL iterations (Fig. \ref{fig:mpo_dielectrics}b), continuously advancing toward the region of high $E_g$ and high $\varepsilon_{\text {total }}$. These results demonstrate that MatInvent can achieve Pareto optimization for two conflicting material properties. Subsequently, more crystal structures were generated by the RL-finetuned diffusion model and ranked according to ML-predicted FoM values, from which 200 structures were selected for DFT validation. As depicted in the Fig. \ref{fig:mpo_dielectrics}c, over 95 \% of the structures exhibit a DFT-calculated $E_g$ exceeding 3.0 eV, which benefits from RL finetuning and the accurate predictive model for $E_g$. In contrast, only about 20 \% of the structures possess a DFT-calculated $\varepsilon_{\text {total }}$ higher than 30, potentially resulting from the poor $\varepsilon_{\text {total }}$ prediction accuracy of ML model (Supplementary Information section \ref{si:sec:dielectric}). Despite the inevitable errors in ML predictions, the DFT property distribution after RL fine-tuning shows a significant shift compared to the pre-trained model (Fig. \ref{fig:mpo_dielectrics}e). Most structures exhibit DFT-calculated FoM exceeding 120, with four structures achieving FoM values greater than 400.
Within a budget of 200 DFT property evaluations, MatInvent successfully identified 17 SUN structures that satisfy all three target criteria (Fig. \ref{fig:mpo_dielectrics}d), highlighting its superior performance in Pareto optimization. All results demonstrate that our RL framework is capable of accomplishing challenging inverse design tasks involving multiple conflicting properties, even when using computationally less expensive rewards with limited accuracy.

\section{Discussion}

MatInvent is a versatile and efficient RL workflow that can tailor the generation of pre-trained diffusion models towards novel material structures with desired properties. This workflow implements policy optimization with reward-weighted KL regularization, experience replay, and diversity filters to ensure efficient optimization and diverse sampling. Across various single-objective design tasks spanning from electronic, magnetic, mechanical, thermal, and physicochemical properties, to synthesizability, MatInvent demonstrates excellent optimization performance, with fast convergence to target values within approximately 60 iterations (1000 property evaluations). Moreover, MatInvent exhibits robust optimization capabilities in design tasks involving multiple conflicting objectives, even with low-precision rewards. Compared to conditional generation approaches that require substantial labeled data for target properties, MatInvent achieves enhanced generative performance in both single- and multi-objective inverse design while dramatically reducing the demand for property assessment.

Despite these strengths, there are several promising directions to enhance MatInvent. The current RL workflow relies on external property evaluators such as ML prediction models and DFT calculations, which may introduce noise and potential biases into the reward signal. Future extensions could incorporate uncertainty-aware or differentiable property predictors to provide informative gradients and enhance learning robustness \cite{uncertainty_rl}. Moreover, real-world materials design tasks frequently involve more than five objectives with varying degrees of importance. Techniques such as curriculum learning \cite{bengio2009curriculum, jeff_curriculum}, Pareto set learning \cite{liu2025pareto}, and preference-conditioned policies \cite{yang2025preference} are worth exploring to enhance MatInvent's performance in multi-objective optimization. Furthermore, material synthesis information could be integrated into the RL framework, such as precursor availability, synthetic route constraints, and synthesizability criteria, which is important for experimental validation and autonomous laboratories.

MatInvent establishes a promising paradigm for inverse material design. Its versatility enables extension to diverse material classes by employing different diffusion models, including perovskites, metal-organic frameworks \cite{mofdiff,mofdiff2}, and two-dimensional materials \cite{2dmat}. The framework can be further adapted to various practical applications, such as catalysis, superconductivity \cite{superconductor}, and quantum computing \cite{okabe2025structural}, using carefully designed RL rewards and property evaluation methods. Moreover, the integration of MatInvent into automated laboratories could offer a compelling avenue for achieving closed-loop material discovery. This general and efficient workflow is poised to attract widespread attention in the material research community.

\clearpage
\section{Methods}

\subsection{Representation of crystal structures}

The periodic structure of crystals arises from the repeating arrangement of atoms in 3D space, and the simplest repeating unit is defined as the unit cell. A unit cell with $N$ atoms can be described by $\mathcal{M}=(\boldsymbol{A}, \boldsymbol{X}, \boldsymbol{L})$, where $\boldsymbol{A}=\left[\boldsymbol{a}_1, \boldsymbol{a}_2, \ldots, \boldsymbol{a}_N\right] \in \mathbb{R}^{h \times N}$ represents the one-hot encoding of atom types, $\boldsymbol{X}=\left[\boldsymbol{x}_1, \boldsymbol{x}_2, \ldots, \boldsymbol{x}_N\right] \in \mathbb{R}^{3 \times N}$ symbolizes atoms’ Cartesian coordinates, and $\boldsymbol{L}=\left[\boldsymbol{l}_1, \boldsymbol{l}_2, \boldsymbol{l}_3\right] \in \mathbb{R}^{3 \times 3}$ expresses the crystal lattice matrix. The volume of a unit cell $V=|\operatorname{det} (\boldsymbol{L})|$ must be non-zero, meaning that $\boldsymbol{L}$ is invertible. Based on periodic boundary conditions, the atomic positions within the unit cell can also be described using fractional coordinates $\boldsymbol{F}=\boldsymbol{L}^{-1} \boldsymbol{X}=\left[\boldsymbol{f}_1, \boldsymbol{f}_2, \ldots, \boldsymbol{f}_N\right] \in [0,1)^{3 \times N}$, which are widely used in crystallography and crystal generation. Thus, a unit cell with $N$ atoms can also be described by $\mathcal{M}=(\boldsymbol{A}, \boldsymbol{F}, \boldsymbol{L})$, and the infinite crystal structure can be represented as
\begin{equation}
\left\{\left(\boldsymbol{a}_i^{\prime}, \boldsymbol{f}_i^{\prime}\right) \mid \boldsymbol{a}_i^{\prime}=\boldsymbol{a}_i, \boldsymbol{f}_i^{\prime}=\boldsymbol{f}_i+\boldsymbol{L} \boldsymbol{k} \mathbf{1}_N, \forall \boldsymbol{k} \in \mathbb{Z}^{3}\right\}
\end{equation}
where elements of $\boldsymbol{k}$ express integer translations of the lattice and $\mathbf{1}$ is a $1 \times n$ matrix of ones to emulate broadcasting.

\subsection{Diffusion models of crystal generation}

This part provides a methodological overview of diffusion models for \textit{de novo} crystal structure generation. The general algorithmic formulation of such models is detailed in Supplementary Information section \ref{si:sec:math_diff}. Implementation details for specific model architectures, including MatterGen \cite{mattergen}, can be found in their original references.

The diffusion models involve two Markov chains: a forward noising process on atom types, atomic fractional coordinates and lattice matrix, and a reverse denoising process learned by a graph neural network (GNN). For the data distribution $q_0$ of 3D crystal structures, $\mathcal{M}_0 \sim q_0\left(\mathcal{M}_0\right)$. The diffusion model approximates $q_0$ with a parameterized ($\theta$) GNN by denoising process in the form of $p_\theta\left(\mathcal{M}_0\right)=\int p_\theta\left(\mathcal{M}_{0: T}\right) d \mathcal{M}_{1: T}$, where $p_\theta\left(\mathcal{M}_{0: T}\right)$ is calculated by
\begin{equation}
p_\theta\left(\mathcal{M}_{0: T}\right)=p_T\left(\mathcal{M}_T\right) \prod_{t=1}^T p_\theta\left(\mathcal{M}_{t-1} \mid \mathcal{M}_t\right),
\end{equation}
and in the timestep $t$ can be described by
\begin{equation}
p_\theta\left(\mathcal{M}_{t-1} \mid \mathcal{M}_t\right)=\mathcal{N}\left(\mu_\theta\left(\mathcal{M}_t, t\right), \sigma^2_t \boldsymbol{I}\right),
\end{equation}
where $\mu_\theta\left(\mathcal{M}_t, t\right)$ is predicted by GNN.

Based on the approximate posterior $q(\mathcal{M}_{1: T} \mid \mathcal{M}_0)$, the denoising process is the reverse of a forward noising process. In the forward process, Gaussian noises are gradually added to $\mathcal{M}$ according to a variance schedule $\beta_1, \ldots, \beta_T$:
\begin{equation}
\begin{split}
q\left(\mathcal{M}_{1: T} \mid \mathcal{M}_0\right)&=\prod_{t=1}^T q\left(\mathcal{M}_t \mid \mathcal{M}_{t-1}\right), \\
\quad q\left(\mathcal{M}_t \mid \mathcal{M}_{t-1}\right)=\mathcal{N} & \left(\sqrt{1-\beta_t} \mathcal{M}_{t-1}, \beta_t \boldsymbol{I}\right),
\end{split}
\end{equation}
Training the diffusion model is conducted by maximizing a variational lower bound on the log-likelihood $\mathbb{E}_q\left[\log p_\theta\left(\mathcal{M}_0\right)\right]$, which is equivalent to optimize the following objective
\begin{equation}
\mathcal{L}(\theta)=\mathbb{E}_{t \sim \mathcal{U}\{0, T\}, \mathcal{M}_t \sim q\left(\mathcal{M}_t \mid \mathcal{M}_0\right)}\left[\left\|\tilde{\boldsymbol{\mu}}\left(\mathcal{M}_0, t\right)-\boldsymbol{\mu}_\theta\left(\mathcal{M}_t, t\right)\right\|^2\right]
\end{equation}
where $\tilde{\boldsymbol{\mu}}$ is the posterior mean of the forward process.

\subsection{Reinforcement learning for crystal diffusion models}

A Markov decision process (MDP) formalizes sequential decision-making problems. It can be characterized by a tuple $(S, A, \rho_0, P, R)$, where $S$ denotes the state space, $A$ represents the action space, $\rho_0$ is the initial state distribution, $P$ specifies the transition kernel, and \( R \) defines the reward function. In every step $t$, the agent observes a state $s_t \in S$, selects an action $a_t \in A$, obtains a reward $R(s_t, a_t)$, and transforms into a subsequent state \( s_{t+1} \sim P(s_{t+1} | s_t, a_t) \). The agent's behavior is determined by its policy $\pi(a | s)$. As the agent interacts with the MDP, it generates trajectories of states and actions $\tau = (s_0, a_0, s_1, a_1, \dots, s_T, a_T)$. The goal of reinforcement learning (RL) is to optimize the agent’s policy $\pi$ to maximize the expected cumulative reward $J_{\mathrm{RL}}(\pi)$ over sampled trajectories:
\begin{equation}
\mathcal{J}_{\mathrm{RL}}(\pi)=\mathbb{E}_{\tau \sim p(\tau \mid \pi)}\left[\sum_{t=0}^T R\left(s_t, a_t\right)\right]
\end{equation}

Our online RL algorithms formulates the denoising process of the diffusion model as a MDP and optimize diffusion models for crystal generation with target properties \cite{dpok, ddpo}. Given a crystal diffusion model $p_\theta (\mathcal{M}_{0: T} )$, parameterized by $\theta$ and the final reward $r(\mathcal{M}_0)$ of crystal $\mathcal{M}_0$ involving single or multiple target material properties, the denoising process can be reframed as a $T$-step MDP:
\begin{equation}
\begin{gathered}
s_t=\mathcal{M}_{T-t}, \quad a_t=\mathcal{M}_{T-t-1}, \\
\rho_0\left(s_0\right)=(\mathcal{N}(0, \boldsymbol{I}), \mathcal{U}(0, 1)), \quad P\left(s_{t+1} \mid s_t, a_t\right)=\delta_{a_t}, \\
\pi \left(a_t \mid s_t\right)=p_\theta\left(\mathcal{M}_{T-t-1} \mid \mathcal{M}_{T-t}\right), \\
R\left(s_t, a_t\right)= \begin{cases}r\left(s_{t+1}\right)=r\left(\mathcal{M}_0\right) & \text { if } t=T-1, \\
0 & \text { otherwise }\end{cases}
\end{gathered}
\end{equation}
where $\delta_y$ is the Dirac delta distribution with nonzero density only at $y$. Sampling the initial state $s_0$ of a trajectory is similar to the first state $\mathcal{M}_T=(\boldsymbol{A}_T, \boldsymbol{F}_T, \boldsymbol{L}_T)$ of the denoising generation, in which $\boldsymbol{A}_T$ and $\boldsymbol{L}_T$ are sampled from $\mathcal{N}(0, \boldsymbol{I})$, and $\boldsymbol{F}_T$ is sampled from $\mathcal{U}(0,1)$. The cumulative reward of every trajectory is
equal to $r\left(\mathcal{M}_0\right)$, because all intermediate rewards are 0, as only the final state $\mathcal{M}_0$ of the denoising process is meaningful for computing crystal properties and rewards. Therefore, optimizing the policy $\pi$ is equivalent to fine-tuning the diffusion model. The common goal in RL fine-tuning of diffusion models is to maximize the expected reward of the generated crystals:
\begin{equation}
\mathcal{J}_{\mathrm{RL}}(\theta) = \mathbb{E}_{p_\theta (\mathcal{M}_{0})}\left[r\left(\mathcal{M}_0\right)\right].
\end{equation}
As depicted in Supplementary Information section \ref{si:sec:math_grad}, the gradient of this objective is
\begin{equation}
\nabla_\theta \mathcal{J}_{\mathrm{RL}}=\mathbb{E}_{p_\theta\left(\mathcal{M}_{0: T}\right)}\left[r\left(\mathcal{M}_0\right) \sum_{t=1}^T \nabla_\theta \log p_\theta\left(\mathcal{M}_{t-1} \mid \mathcal{M}_t\right)\right].
\end{equation}
The risk of fine-tuning solely based on rewards related to target properties is that the diffusion model may overfit to the rewards and move too far away from the initial state (pre-trained model) \cite{dpok}. To retain the broad material knowledge that the diffusion model has learned from the pre-training dataset for generating reasonable and valid crystal structures, we add the reward-weighted KL between the pre-trained and current fine-tuned models as a regularizer to the objective function according to:
\begin{equation}
\mathbb{E}_{p_\theta\left(\mathcal{M}_{0: T}\right)}
\left[
(\lambda - r\left(\mathcal{M}_0\right)) \sum_{t=1}^T \operatorname{KL}\left(p_\theta\left(\mathcal{M}_{t-1} \mid \mathcal{M}_t\right) \| p_{\mathrm{pre}}\left(\mathcal{M}_{t-1} \mid \mathcal{M}_t\right)\right)
\right],
\end{equation}
where $\lambda$ is a constant slightly larger than the maximum reward and more details are in Supplementary Information section \ref{si:sec:math_kl}. The reward weight allows the current diffusion model to appropriately move away from the pre-trained model \cite{reinvent}, thereby encouraging the model to shift its distribution to higher reward regions. Thus, the final gradient to optimize the  RL objective is:
\begin{equation}
\begin{split}
& - \alpha r\left(\mathcal{M}_0\right) \sum_{t=1}^T \nabla_\theta \log p_\theta\left(\mathcal{M}_{t-1} \mid \mathcal{M}_t\right) \\
& + \beta (\lambda - r\left(\mathcal{M}_0\right)) \sum_{t=1}^T \nabla_\theta \operatorname{KL}\left(p_\theta\left(\mathcal{M}_{t-1} \mid \mathcal{M}_t\right) \| p_{\mathrm{pre}}\left(\mathcal{M}_{t-1} \mid \mathcal{M}_t\right)\right)
\end{split}
\end{equation}
where $\alpha$ and $\beta$ are the weights of reward and KL regularization, respectively.

\paragraph{Experience replay} The experience replay \cite{experience-replay, augmented-memory} is integrated into MatInvent, which is used to improve the stability and efficiency of RL by storing past high-reward crystals and reusing them during model fine-tuning. It breaks the correlation between consecutive experiences by sampling from a buffer of previous experiences (called the replay buffer) rather than relying only on the most recent experience. Specifically, the size of replay buffer is set to 100. When the number of stored crystal structures exceeded this capacity, only the 100 structures with the highest rewards are retained. In each RL iteration, 10 crystal structures are randomly sampled from the replay buffer and combined with the top 50 \% rewarded structures from the current iteration for model fine-tuning. After fine-tuning, the top 50 \% rewarded structures are added to the replay buffer, applying a deduplication criterion whereby only the highest-rewarded structure is preserved for each unique chemical composition.

\paragraph{Diversity filter (DF)} We draw inspiration for DFs from \cite{diversity-filter, thomas2022augmented} with small modifications. In this work. DFs linearly penalize crystals with non-unique  chemical compositions based on the number of previous occurrences, which acts as a more lenient version of the unique DF, i.e., directly truncate the reward to 0 \cite{diversity-filter}. The reward is transformed according to the number of previous occurrences (Occ) beyond an allowed tolerance (Tol) until a hard threshold is reached, referred to as the buffer (Buff):
\begin{equation}
\text { Filtered reward }=\left\{\begin{array}{cl}
r\left(\mathcal{M}_0\right) \times \frac{\mathrm{Occ}- \text { Tol }}{\text { Buff }- \text { Tol }} & \text { if } \mathrm{Tol}<\mathrm{Occ}<\text { Buff } \\[3pt]
r\left(\mathcal{M}_0\right) & \text { if } \quad \text { Occ } \leq \text { Tol } \\[3pt]
0 & \text { if } \quad \text { Occ } \geq \text { Buff }
\end{array}\right.
\end{equation}
where Tol is set to 3 and Buff is set to 6. Selective memory purge will be triggered for material structures that remain in the replay buffer but are penalized by the diversity filter, resulting in their removal from the replay buffer \cite{augmented-memory}. That is, crystals with the same chemical composition as previously generated samples are assigned reduced rewards and subsequently removed from the replay buffer.

\clearpage
\section*{Data availability}
The diffusion models were pre-trained on the open-source Alex-MP \cite{mattergen} or MP-20 \cite{mp,cdvae} datasets. Checkpoint files for the diffusion model and property prediction model are available at Hugging Face \url{https://huggingface.co/jwchen25/MatInvent}.

\section*{Code availability}
The source code for MatInvent is available at GitHub \url{https://github.com/schwallergroup/matinvent}.

\section*{Acknowledgments}
J.C., J.G., E.F. and P.S. acknowledge support from the NCCR Catalysis (grant number 225147), a National Centre of Competence in Research funded by the Swiss National Science Foundation. J.G. (PGSD-521528389) acknowledges support from the Natural Sciences and Engineering Research Council of Canada (NSERC).

\section*{Author contributions}
J.C. contributed to methodology, model development, coding, writing, visualization, and assessment. J.G. and E.F. contributed to methodology, model design and writing. P.S. contributed to conceptualization, methodology, model design, writing, assessment, funding and project supervision.

\section*{Competing interests}
The authors declare no competing interests.


\clearpage
\bibliographystyle{unsrt}
\bibliography{main_ref}

\clearpage
\startappendix




\renewcommand{\thesection}{\Alph{section}}
\renewcommand{\thepage}{S\arabic{page}}
\renewcommand{\thetable}{S\arabic{table}}
\renewcommand{\thefigure}{S\arabic{figure}}
\renewcommand{\theequation}{S\arabic{equation}}

\setcounter{section}{0}
\setcounter{equation}{0}
\setcounter{figure}{0}
\setcounter{table}{0}
\setcounter{page}{1}

\section*{Supplementary Information}

\addcontentsline{toc}{section}{Supplementary Information}
\subsection*{Contents}
\startcontents[SI]
\printcontents[SI]{}{1}{\setcounter{tocdepth}{3}}
\stopcontents[SI]
\resumecontents[SI]

\clearpage
\section{Algorithm details and mathematical proofs}
\subsection{Diffusion models of crystal generation} \label{si:sec:math_diff}
This part provides the general algorithmic formulation of diffusion models for \textit{de novo} crystal structure generation. These diffusion models involve two Markov chains: a forward noising process on atom types, atomic fractional coordinates and lattice matrix, and a reverse denoising process learned by a graph neural network.

 A unit cell of crystal with $N$ atoms is described by $\mathcal{M}=(\boldsymbol{A}, \boldsymbol{X}, \boldsymbol{L})$, where $\boldsymbol{A}=\left[\boldsymbol{a}_1, \boldsymbol{a}_2, \ldots, \boldsymbol{a}_N\right] \in \mathbb{R}^{h \times N}$ represents the one-hot encoding of atom types, $\boldsymbol{X}=\left[\boldsymbol{x}_1, \boldsymbol{x}_2, \ldots, \boldsymbol{x}_N\right] \in \mathbb{R}^{3 \times N}$ symbolizes atoms’ Cartesian coordinates, and $\boldsymbol{L}=\left[\boldsymbol{l}_1, \boldsymbol{l}_2, \boldsymbol{l}_3\right] \in \mathbb{R}^{3 \times 3}$ expresses the crystal lattice matrix. Based on periodic boundary conditions, the atomic positions within the unit cell can also be described using fractional coordinates $\boldsymbol{F}=\boldsymbol{L}^{-1} \boldsymbol{X}=\left[\boldsymbol{f}_1, \boldsymbol{f}_2, \ldots, \boldsymbol{f}_N\right] \in [0,1)^{3 \times N}$, which are widely used in crystallography and crystal generation. Thus, a unit cell with $N$ atoms can also be described by $\mathcal{M}=(\boldsymbol{A}, \boldsymbol{F}, \boldsymbol{L})$.

 \paragraph{Diffusion on lattice $\boldsymbol{L}$}
The diffusion on the continuous variable $\boldsymbol{L}$ is based on Denoising Diffusion Probabilistic
Model (DDPM) \citeSI{si_ho2020denoising}. Specifically, in the forward process, Gaussian noises are gradually added to $\boldsymbol{L}$ according to a variance schedule $\beta_1, \ldots, \beta_T$:
\begin{equation} \label{eq2}
\begin{split}
q\left(\boldsymbol{L}_{1: T} \mid \boldsymbol{L}_0\right)&=\prod_{t=1}^T q\left(\boldsymbol{L}_t \mid \boldsymbol{L}_{t-1}\right), \\
\quad q\left(\boldsymbol{L}_t \mid \boldsymbol{L}_{t-1}\right)=\mathcal{N} & \left(\boldsymbol{L}_t \mid \sqrt{1-\beta_t} \boldsymbol{L}_{t-1}, \beta_t \boldsymbol{I}\right),
\end{split}
\end{equation}
which can be expressed as the probability conditional on the initial state:
\begin{equation} \label{eq3}
q\left(\boldsymbol{L}_t \mid \boldsymbol{L}_0\right)=\mathcal{N}\left(\boldsymbol{L}_t \mid \sqrt{\bar{\alpha}_t} \boldsymbol{L}_0,\left(1-\bar{\alpha}_t\right) \boldsymbol{I}\right),
\end{equation}
using $\alpha_t=1-\beta_t$ and $\bar{\alpha}_t=\prod_{s=1}^t \alpha_s$.

The reverse process is defined by:
\begin{equation}
\begin{split}
p_\theta\left(\boldsymbol{L}_{0: T}\right)=& p\left(\boldsymbol{L}_T \right) \prod_{t=1}^T p_\theta\left(\boldsymbol{L}_{t-1} \mid \boldsymbol{L}_t\right), \\
p_\theta\left(\boldsymbol{L}_{t-1} \mid \boldsymbol{L}_t\right)=&\mathcal{N} \left(\boldsymbol{L}_{t-1}\mid \boldsymbol{\mu}_{\theta, \boldsymbol{L}}\left(\mathcal{M}_t, t\right), \sigma^2_t \boldsymbol{I}\right),
\end{split}
\end{equation}
where $\boldsymbol{\mu}_{\theta, \boldsymbol{L}}\left(\mathcal{M}_t, t\right)=\frac{1}{\sqrt{\alpha_t}}\left(\boldsymbol{L}_t-\frac{\beta_t}{\sqrt{1-\bar{\alpha}_t}} \hat{\boldsymbol{\epsilon}}_{\theta, \boldsymbol{L}}\left(\mathcal{M}_t, t\right)\right)$ and $p\left(\boldsymbol{L}_T\right)=\mathcal{N}(0, \boldsymbol{I})$. The denoising term $\hat{\boldsymbol{\epsilon}}_{\theta, \boldsymbol{L}}\left(\mathcal{M}_t, t\right) \in \mathbb{R}^{3 \times 3}$
 is predicted by the equivariant graph neural network $\theta\left(\mathcal{M}_t, t\right) = \theta\left(\boldsymbol{L}_t, \boldsymbol{F}_t, \boldsymbol{A}_t, t\right)$.0
 
 For training the denoising model $\theta$, let $\boldsymbol{L}_t=\sqrt{\bar{\alpha}_t} \boldsymbol{L}_0+\sqrt{1-\bar{\alpha}_t} \boldsymbol{\epsilon}_{\boldsymbol{L}}$ and $\boldsymbol{\epsilon}_{\boldsymbol{L}} \sim \mathcal{N}(0, \boldsymbol{I})$ according to Eq. (\ref{eq3}). The training objective is denoted as the $\ell_2$ loss between $\boldsymbol{\epsilon}_{\boldsymbol{L}}$ and $\hat{\boldsymbol{\epsilon}}_{\theta, \boldsymbol{L}}$:
 \begin{equation} \label{eq5}
\mathcal{L}_{\boldsymbol{L}}=\mathbb{E}_{t \sim \mathcal{U}(1, T)} \left[\left\|\boldsymbol{\epsilon}_{\boldsymbol{L}}-\hat{\boldsymbol{\epsilon}}_{\theta,\boldsymbol{L}}\left(\mathcal{M}_t, t\right)\right\|^2\right] .
\end{equation}

\paragraph{Diffusion on atom types $\boldsymbol{A}$}
The discrete atom types $\boldsymbol{A}$ can be simply considered as continuous variables in real space $\mathbb{R}^{h \times N}$, facilitating the DDPM-based approach for diffusion on atom types, as also shown in \citeSI{si_edm}. Similar to diffusion on $\boldsymbol{L}$ (Eq. \ref{eq2}-\ref{eq5}), the forward process of $\boldsymbol{A}$ is denoted as
\begin{equation}
q\left(\boldsymbol{A}_t \mid \boldsymbol{A}_0\right)=\mathcal{N}\left(\boldsymbol{A}_t \mid \sqrt{\bar{\alpha}_t} \boldsymbol{A}_0,\left(1-\bar{\alpha}_t\right) \boldsymbol{I}\right),
\end{equation}
the reverse process is expressed as
\begin{equation}
p_\theta\left(\boldsymbol{A}_{t-1} \mid \boldsymbol{A}_t\right)=\mathcal{N} \left(\boldsymbol{A}_{t-1}\mid \boldsymbol{\mu}_{\theta, \boldsymbol{A}}\left(\mathcal{M}_t, t\right), \sigma^2_t \boldsymbol{I}\right),
\end{equation}
and the training objective for diffusion on $\boldsymbol{A}$ is
\begin{equation}
\mathcal{L}_{\boldsymbol{A}}=\mathbb{E}_{t \sim \mathcal{U}(1, T)} \left[\left\|\boldsymbol{\epsilon}_{\boldsymbol{A}}-\hat{\boldsymbol{\epsilon}}_{\theta, \boldsymbol{A}}\left(\mathcal{M}_t, t\right)\right\|^2\right] .
\end{equation}

\paragraph{Diffusion on atom positions $\boldsymbol{F}$}
As the domain of fractional coordinates $[0,1)^{3 \times N}$ forms a quotient space $\mathbb{R}^{3 \times N} / \mathbb{Z}^{3 \times N}$, the score matching method \citeSI{si_song2020score} with wrapped normal distribution \citeSI{si_bortoli2022riemannian} is used to achieve diffusion on $\boldsymbol{F}$ \citeSI{si_diffcsp}. The forward process is implemented by wrapped normal distribution to maintain periodic translation invariance according to:
\begin{equation}
q\left(\boldsymbol{F}_t \mid \boldsymbol{F}_0\right) = \mathcal{N}_W\left(\boldsymbol{F}_t \mid \boldsymbol{F}_0, \sigma_t^2 \boldsymbol{I}\right), \ \ 
\boldsymbol{F}_t=w \left(\boldsymbol{F}_0+\sigma_t \boldsymbol{\epsilon}_{\boldsymbol{F}}\right),
\end{equation}
where $\boldsymbol{\epsilon}_{\boldsymbol{F}} \sim \mathcal{N}(0, \boldsymbol{I})$ and $w(\cdot)$ retains the fractional part of the input. The noise scale $\sigma_t$ obeys the exponential scheduler:
$\sigma_0=0$ and $\sigma_t=\sigma_1\left(\frac{\sigma_T}{\sigma_1}\right)^{\frac{t-1}{T-1}}$, if $t>0$.

For the reverse process, $\boldsymbol{F}_T \sim \mathcal{U}(0,1)$ and $\boldsymbol{F}_0$ are generated using a two-step predictor-corrector sampler method \citeSI{si_song2020score, si_diffcsp} with the denoising term $\hat{\boldsymbol{\epsilon}}_{\theta, \boldsymbol{F}}\left(\mathcal{M}_t, t\right) \in \mathbb{R}^{3 \times N}$:
\begin{equation}
p_{\theta}\left(\boldsymbol{F}_{t-1} \mid \mathcal{M}_t\right) = p_P\left(\left.\boldsymbol{F}_{t-\frac{1}{2}} \right\rvert\, \boldsymbol{L}_t, \boldsymbol{F}_t, \boldsymbol{A}_t\right)
p_C \left( \boldsymbol{F}_{t-1} \mid \boldsymbol{L}_{t-1}, \boldsymbol{F}_{t-\frac{1}{2}}, \boldsymbol{A}_{t-1}\right),
\end{equation}
where $p_P, p_C$ are the transitions of the predictor and corrector.

The training objective from score matching of $\boldsymbol{F}$ is
\begin{equation}
\mathcal{L}_{\boldsymbol{F}}=\mathbb{E}_{t \sim \mathcal{U}(1, T)}\left[\lambda_t\left\|\nabla \log q\left(\boldsymbol{F}_t \mid \boldsymbol{F}_0\right)-\hat{\boldsymbol{\epsilon}}_{\theta, \boldsymbol{F}}\left(\mathcal{M}_t, t\right)\right\|^2\right]
\end{equation}
where $\lambda_t=\mathbb{E}_{\boldsymbol{F}_t}^{-1}\left[\left\|\nabla \log q\left(\boldsymbol{F}_t \mid \boldsymbol{F}_0\right)\right\|^2\right]$ is calculated by Monte-Carlo sampling.

\clearpage
\subsection{Gradient of RL objective function} \label{si:sec:math_grad}
A common goal in RL fine-tuning of diffusion models is to maximize the expected reward of the generated crystal structures:
\begin{equation}
\min _\theta \mathbb{E}_{p_\theta\left(\mathcal{M}_0 \right)}\left[-r\left(\mathcal{M}_0\right)\right]
\end{equation}
The gradient of this objective function can be obtained as follows:
\begin{equation}
\begin{aligned}
& \nabla_\theta \mathbb{E}_{p_\theta\left(\mathcal{M}_0 \right)}\left[-r\left(\mathcal{M}_0\right)\right]=\nabla_\theta \int p_\theta\left(\mathcal{M}_0 \right) r\left(\mathcal{M}_0\right) d \mathcal{M}_0 \\
= & -\nabla_\theta \int\left(\int p_\theta\left(\mathcal{M}_{0: T} \right) d \mathcal{M}_{1: T}\right) r\left(\mathcal{M}_0\right) d \mathcal{M}_0 \\
= & -\int \nabla_\theta \log p_\theta\left(\mathcal{M}_{0: T} \right) \times r\left(\mathcal{M}_0\right) \times p_\theta\left(\mathcal{M}_{0: T} \right) d \mathcal{M}_{0: T} \\
= & -\int \nabla_\theta \log \left(p_T\left(\mathcal{M}_T \right) \prod_{t=1}^T p_\theta\left(\mathcal{M}_{t-1} \mid \mathcal{M}_t\right)\right) \times r\left(\mathcal{M}_0\right) \times p_\theta\left(\mathcal{M}_{0: T} \right) d \mathcal{M}_{0: T} \\
= & \mathbb{E}_{p_\theta\left(\mathcal{M}_{0: T} \right)}\left[- r\left(\mathcal{M}_0\right) \sum_{t=1}^T \nabla_\theta \log p_\theta\left(\mathcal{M}_{t-1} \mid \mathcal{M}_t\right)\right].
\end{aligned}
\end{equation}

\clearpage
\subsection{KL regularization} \label{si:sec:math_kl}
To prevent overfitting to task-specific rewards while preserving material knowledge that the diffusion model has learned from the pre-training dataset for generating reasonable and valid crystal structures, we augment the RL objective function with a reward-weighted KL divergence regularizer between the pre-trained and fine-tuned diffusion models. Unlike the language models in which the KL regularizer is computed over the entire sequence/trajectory (of tokens), in diffusion models, it makes sense to compute it only
for the final crystal structures $\operatorname{KL}\left(p_\theta\left(\mathcal{M}_0 \right) \| p_{\text{pre }}\left(\mathcal{M}_0 \right)\right)$. Unfortunately, $p_\theta(\mathcal{M}_0 )$ is a marginal and its closed-form is unknown. Thus, it is converted to an upper-bound format. From data processing inequality with the Markov kernel, we have
\begin{equation}
\left.\operatorname{KL}\left(p_\theta\left(\mathcal{M}_0 \right)\right) \| p_{\text {pre }}\left(\mathcal{M}_0 \right)\right) \leq \operatorname{KL}\left(p_\theta\left(\mathcal{M}_{0: T} \right) \| p_{\text {pre }}\left(\mathcal{M}_{0: T} \right)\right)
\end{equation}
where a periodic crystal is described by $\mathcal{M}=(\boldsymbol{A}, \boldsymbol{X}, \boldsymbol{L})$. Using the Markov property of $p_\theta$ and $p_{\text{pre}}$, it can be converted into
\begin{equation}
\begin{aligned}
& \operatorname{KL}\left(p_\theta\left(\mathcal{M}_{0: T} \right) \| p_{\text {pre }}\left(\mathcal{M}_{0: T} \right)\right)=\int p_\theta\left(\mathcal{M}_{0: T} \right) \times \log \frac{p_\theta\left(\mathcal{M}_{0: T} \right)}{p_{\text {pre }}\left(\mathcal{M}_{0: T} \right)} d \mathcal{M}_{0: T} \\
= & \int p_\theta\left(\mathcal{M}_{0: T} \right) \log \frac{p_\theta\left(\mathcal{M}_T \right) \prod_{t=1}^T p_\theta\left(\mathcal{M}_{t-1} \mid \mathcal{M}_t\right)}{p_{\text {pre }}\left(\mathcal{M}_T \right) \prod_{t=1}^T p_{\text {pre }}\left(\mathcal{M}_{t-1} \mid \mathcal{M}_t\right)} d \mathcal{M}_{0: T} \\
= & \int p_\theta\left(\mathcal{M}_{0: T} \right)\left(\log \frac{p_\theta\left(\mathcal{M}_T \right)}{p_{\text {pre }}\left(\mathcal{M}_T \right)}+\sum_{t=1}^T \log \frac{p_\theta\left(\mathcal{M}_{t-1} \mid \mathcal{M}_t\right)}{p_{\text {pre }}\left(\mathcal{M}_{t-1} \mid \mathcal{M}_t\right)}\right) d \mathcal{M}_{0: T} \\
= & \mathbb{E}_{p_\theta\left(\mathcal{M}_{0: T} \right)}\left[\sum_{t=1}^T \log \frac{p_\theta\left(\mathcal{M}_{t-1} \mid \mathcal{M}_t\right)}{p_{\text {pre }}\left(\mathcal{M}_{t-1} \mid \mathcal{M}_t\right)}\right]=\sum_{t=1}^T \mathbb{E}_{p_\theta\left(\mathcal{M}_{t: T} \right)} \mathbb{E}_{p_\theta\left(\mathcal{M}_{0: t-1} \mid \mathcal{M}_{t: T}\right)}\left[\log \frac{p_\theta\left(\mathcal{M}_{t-1} \mid \mathcal{M}_t\right)}{p_{\text {pre }}\left(\mathcal{M}_{t-1} \mid \mathcal{M}_t\right)}\right] \\
= & \sum_{t=1}^T \mathbb{E}_{p_\theta\left(\mathcal{M}_t \right)} \mathbb{E}_{p_\theta\left(\mathcal{M}_{0: t-1} \mid \mathcal{M}_t\right)}\left[\log \frac{p_\theta\left(\mathcal{M}_{t-1} \mid \mathcal{M}_t\right)}{p_{\text {pre }}\left(\mathcal{M}_{t-1} \mid \mathcal{M}_t\right)}\right]=\sum_{t=1}^T \mathbb{E}_{p_\theta\left(\mathcal{M}_t \right)} \mathbb{E}_{p_\theta\left(\mathcal{M}_{t-1} \mid \mathcal{M}_t\right)}\left[\log \frac{p_\theta\left(\mathcal{M}_{t-1} \mid \mathcal{M}_t\right)}{p_{\text {pre }}\left(\mathcal{M}_{t-1} \mid \mathcal{M}_t\right)}\right] \\
= & \sum_{t=1}^T \mathbb{E}_{p_\theta\left(\mathcal{M}_t \right)}\left[\operatorname{KL}\left(p_\theta\left(\mathcal{M}_{t-1} \mid \mathcal{M}_t\right) \| p_{\text {pre }}\left(\mathcal{M}_{t-1} \mid \mathcal{M}_t\right)\right)\right] .
\end{aligned}
\end{equation}
Finally,
\begin{equation} \label{equ:rl}
\left.\operatorname{KL}\left(p_\theta\left(\mathcal{M}_0 \right)\right) \| p_{\text {pre }}\left(\mathcal{M}_0 \right)\right) \leqslant \sum_{t=1}^T \mathbb{E}_{p_\theta\left(\mathcal{M}_t \right)}\left[\operatorname{KL}\left(p_\theta\left(\mathcal{M}_{t-1} \mid \mathcal{M}_t\right) \| p_{\text {pre }}\left(\mathcal{M}_{t-1} \mid \mathcal{M}_t \right)\right)\right] .
\end{equation}
For online fine-tuning, we need to regularize $\sum_{t=1}^T \mathbb{E}_{p_\theta\left(\mathcal{M}_t \right)}\left[\operatorname{KL}\left(p_\theta\left(\mathcal{M}_{t-1} \mid \mathcal{M}_t\right) \| p_{\text {pre }}\left(\mathcal{M}_{t-1} \mid \mathcal{M}_t \right)\right)\right]$.
By the product rule, we can have the gradient of objective Eq. \ref{equ:rl}
\begin{equation}
\begin{aligned}
\nabla_\theta & \sum_{t=1}^T \mathbb{E}_{p_\theta\left(\mathcal{M}_t\right)}\left[\operatorname{KL}\left(p_\theta\left(\mathcal{M}_{t-1} \mid \mathcal{M}_t\right) \| p_{\text {pre }}\left(\mathcal{M}_{t-1} \mid \mathcal{M}_t\right)\right)\right] \\
& =\sum_{t=1}^T \mathbb{E}_{p_\theta\left(\mathcal{M}_t\right)}\left[\nabla_\theta \operatorname{KL}\left(p_\theta\left(\mathcal{M}_{t-1} \mid \mathcal{M}_t\right) \| p_{\text {pre }}\left(\mathcal{M}_{t-1} \mid \mathcal{M}_t\right)\right)\right] \\
& +\sum_{t=1}^T \mathbb{E}_{p_\theta\left(\mathcal{M}_t\right)}\left[\sum_{t^{\prime}>t}^T \nabla_\theta \log p_\theta\left(\mathcal{M}_{t^{\prime}-1} \mid \mathcal{M}_{t^{\prime}}\right) \cdot \operatorname{KL}\left(p_\theta\left(\mathcal{M}_{t-1} \mid \mathcal{M}_t\right) \| p_{\text {pre }}\left(\mathcal{M}_{t-1} \mid \mathcal{M}_t\right)\right)\right],
\end{aligned}
\end{equation}
which treats the sum of conditional KL-divergences along the future trajectory as a scalar reward at each step. However, computing these sums is more inefficient than just the first term. Empirically, we find that regularizing only the first term already works well, so that
\begin{equation}
\begin{aligned}
& \nabla_\theta \sum_{t=1}^T \mathbb{E}_{p_\theta\left(\mathcal{M}_t\right)}\left[\operatorname{KL}\left(p_\theta\left(\mathcal{M}_{t-1} \mid \mathcal{M}_t\right) \| p_{\text {pre }}\left(\mathcal{M}_{t-1} \mid \mathcal{M}_t\right)\right)\right] \\
& \quad \approx \sum_{t=1}^T \mathbb{E}_{p_\theta\left(\mathcal{M}_t\right)}\left[\nabla_\theta \operatorname{KL}\left(p_\theta\left(\mathcal{M}_{t-1} \mid \mathcal{M}_t\right) \| p_{\text {pre }}\left(\mathcal{M}_{t-1} \mid \mathcal{M}_t\right)\right)\right] \\
& \quad \approx \mathbb{E}_{p_\theta\left(\mathcal{M}_{0: T} \right)}\left[ \sum_{t=1}^T \nabla_\theta \operatorname{KL}\left(p_\theta\left(\mathcal{M}_{t-1} \mid \mathcal{M}_t\right) \| p_{\text {pre }}\left(\mathcal{M}_{t-1} \mid \mathcal{M}_t\right)\right)\right]
\end{aligned}
\end{equation}
And the corresponding
\begin{equation}
\begin{aligned}
& \sum_{t=1}^T \mathbb{E}_{p_\theta\left(\mathcal{M}_t\right)}\left[\operatorname{KL}\left(p_\theta\left(\mathcal{M}_{t-1} \mid \mathcal{M}_t\right) \| p_{\text {pre }}\left(\mathcal{M}_{t-1} \mid \mathcal{M}_t\right)\right)\right] \\
& \quad \approx \mathbb{E}_{p_\theta\left(\mathcal{M}_{0: T} \right)}\left[ \sum_{t=1}^T \operatorname{KL}\left(p_\theta\left(\mathcal{M}_{t-1} \mid \mathcal{M}_t\right) \| p_{\text {pre }}\left(\mathcal{M}_{t-1} \mid \mathcal{M}_t\right)\right)\right]
\end{aligned}
\end{equation}
And the reward-weighted KL regularization between the pre-trained and current fine-tuned models is defined by
\begin{equation}
\mathbb{E}_{p_\theta\left(\mathcal{M}_{0: T}\right)}
\left[
(\lambda - r\left(\mathcal{M}_0\right)) \sum_{t=1}^T \operatorname{KL}\left(p_\theta\left(\mathcal{M}_{t-1} \mid \mathcal{M}_t\right) \| p_{\mathrm{pre}}\left(\mathcal{M}_{t-1} \mid \mathcal{M}_t\right)\right)
\right],
\end{equation}
where $\lambda$ is a constant slightly larger than the maximum reward. And corresponding gradient:
\begin{equation}
\mathbb{E}_{p_\theta\left(\mathcal{M}_{0: T}\right)}
\left[
(\lambda - r\left(\mathcal{M}_0\right)) \sum_{t=1}^T \nabla_\theta \operatorname{KL}\left(p_\theta\left(\mathcal{M}_{t-1} \mid \mathcal{M}_t\right) \| p_{\mathrm{pre}}\left(\mathcal{M}_{t-1} \mid \mathcal{M}_t\right)\right)
\right],
\end{equation}

\clearpage
\subsection{Implementation details}
The diffusion models were built with PyTorch Geometric \citeSI{si_Lenssen2019} and PyTorch \citeSI{si_paszke2019pytorch}. The crystal structures were processed using the Atomic Simulation Environment (ASE) package \citeSI{si_larsen2017atomic} and pymatgen \citeSI{si_pymatgen}. Matplotlib \citeSI{si_hunter2007matplotlib} was used to draw the graphs presented in this work. During the RL process, the diffusion model was fine-tuned on a single NVIDIA H100 GPU at float32 precision.

\clearpage
\section{Analysis of generated structures} \label{si:sec:metrics}

\subsection{Stability, uniqueness and novelty (SUN)}
The thermodynamically Stability ($E_{hull} <$ 0.1 eV/atom), Uniqueness, and Novelty (SUN) \citeSI{si_mattergen} of each generated structure was evaluated by MatterGen's method using their Alex-MP reference dataset and code \citeSI{si_mattergen}. For DFT evaluation, $E_{hull}$ was calculated by DFT energy after relaxation. For SUN ratios of different models, $E_{hull}$ was obtained by MLIP energy after geometry optimization using MatterSim \citeSI{si_mattersim} due to the high computational cost. To evaluate and compare SUN ratios across different models, each diffusion model generated 1,024 structures, with the SUN ratio defined as the percentage of structures satisfying the SUN criteria.

\subsection{Diversity ratio}
The uniqueness index in the SUN metric reflects the ratio of unique structures in the generated crystals. However, in experimental studies, chemical composition is also an important focus. Furthermore, the unique composition ratio is often smaller than the unique structure ratio, because structures with different compositions must have different crystal structures.

During RL fine-tuning, the diffusion generative model tends to produce crystals in specific regions with high rewards, leading to reduced sample diversity. Therefore, the diversity ratio metric is defined as the ratio between the number of unique chemical compositions ($u$) and the number of all crystal structures ($s$) generated during the RL process:
\begin{equation}
\text{Diversity ratio}=\frac{u}{s},
\end{equation}

\subsection{Visualization}
The generated crystal structures were visualized using Crystal Toolkit \citeSI{si_crystal_toolkit} with the default setting. A uniform atomic radius 0.5 $\text{\AA}$ was used, while CrystalNN bonding algorithm \citeSI{si_crystalnn} was used for chemical bonds. All 3D visualization images show atoms, bonds, unit cell and polyhedra.

\clearpage
\section{Generalizability and ablation study of MatInvent}

\subsection{Generalizability of MatInvent on different diffusion models} \label{si:sec:general}

\begin{figure}[htp]
\centering
\includegraphics[width=1.0\textwidth]{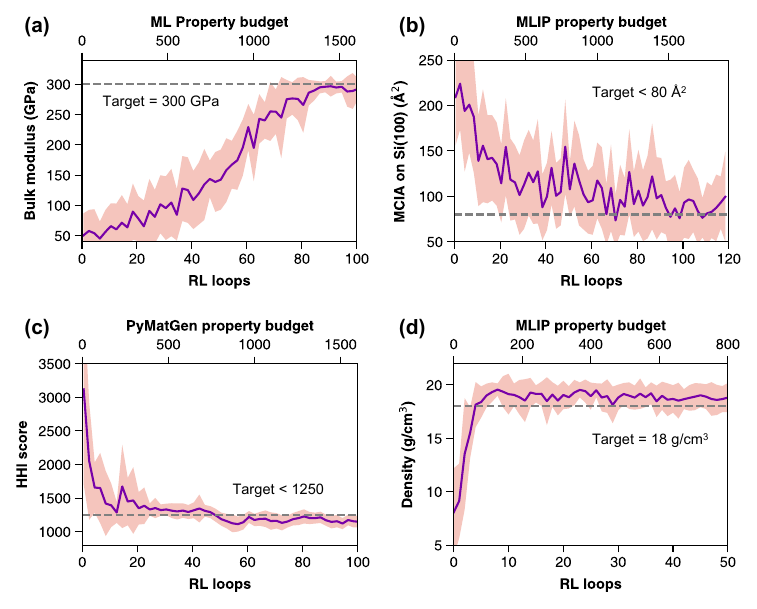}
\caption{
The optimization curves of MatInvent workflow using DiffCSP diffusion model on different inverse design tasks with a single target property:
(a) bulk modulus of 300 GPa;
(b) minimal co-incident area (MCIA) below 80 $\text{\AA}^{2}$;
(c) Herfindahl–Hirschman index (HHI) score below 1250;
and (d) density of 18.0 g/cm$^3$.
The curves represents the average values of the target properties of the generated structures in each RL iteration, while the shading depicts standard deviation.
}
\label{fig:diffcsp}
\end{figure}

\begin{figure}[htp]
\centering
\includegraphics[width=1.0\textwidth]{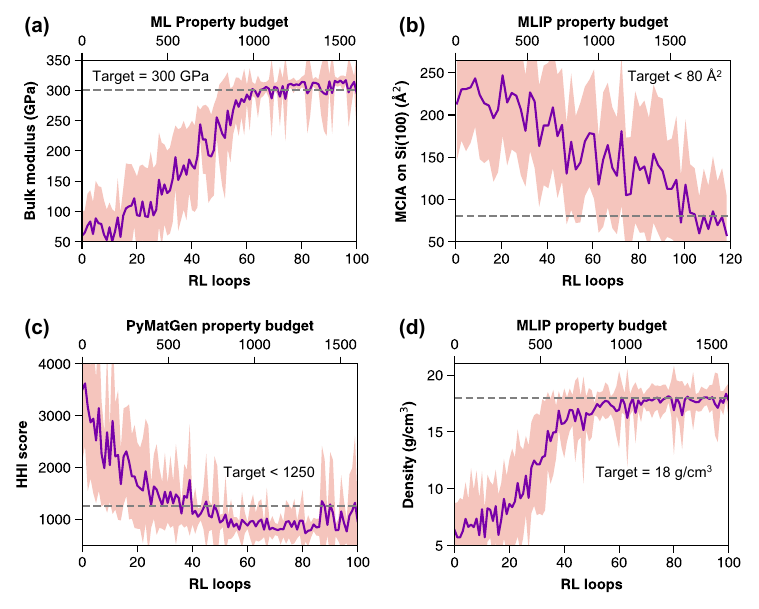}
\caption{
The optimization curves of MatInvent workflow using EquiCSP diffusion model on different inverse design tasks with a single target property:
(a) bulk modulus of 300 GPa;
(b) minimal co-incident area (MCIA) below 80 $\text{\AA}^{2}$;
(c) Herfindahl–Hirschman index (HHI) score below 1250;
and (d) density of 18.0 g/cm$^3$.
The curves represents the average values of the target properties of the generated structures in each RL iteration, while the shading depicts standard deviation.
}
\label{fig:equicsp}
\end{figure}

As illustrated in Fig. \ref{fig:diffcsp} and \ref{fig:equicsp}, we evaluted MatInvent across two different diffusion models, DiffCSP \citeSI{si_diffcsp} and EquiCSP \citeSI{si_linequivariant}, on four single-property optimization tasks: (1) bulk modulus of 300 GPa;
(2) MCIA below 80 $\text{\AA}^{2}$;
(3) HHI score below 1250;
and (d) density of 18.0 g/cm$^3$. The results demonstrate that MatInvent iteratively optimizes the diffusion models through RL, progressively driving the mean value of the target property of generated structures toward the optimization objective. For most tasks, MatInvent achieves rough convergence within 60 iterations. Thus, MatInvent is a general-purpose RL workflow that is compatible with different diffusion model architectures.

\clearpage
\subsection{Ablation study} \label{si:sec:ablation}
\subsubsection{Effect of geometry optimization and filter} \label{si:sec:ablation:optf}
\begin{figure}[htp]
\centering
\includegraphics[width=1.0\textwidth]{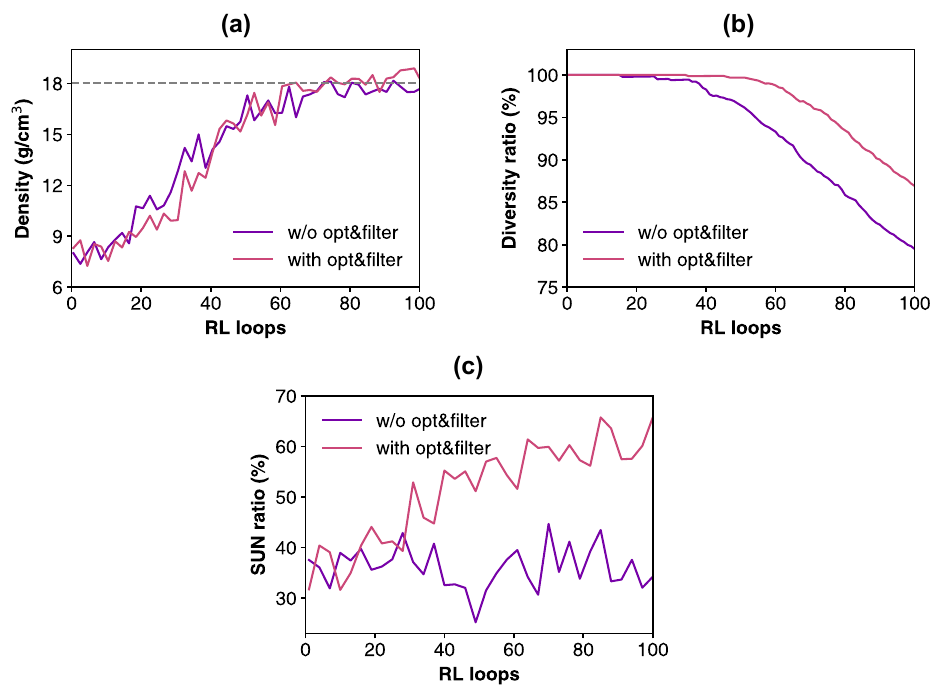}
\caption{
The RL optimization curves (a), composition diversity ratios (b), and SUN ratios (c) of generated structures during the MatInvent process with or without MLIP-based geometry optimization (opt) and SUN filter prior to property evaluation, for the target density of 18.0 g/cm$^3$.
The optimization curves represents the average values of density of the generated structures in each RL iteration.
}
\label{fig:ablation_optf}
\end{figure}

As shown in Fig. \ref{fig:ablation_optf}a, MLIP-based geometric optimization (opt) and SUN filtering prior to property assessment exert negligible influence on RL optimization efficiency. However, Fig. \ref{fig:ablation_optf}b and c reveal that structure optimization and SUN filtering substantially enhance the compositional diversity and SUN ratio of structures generated by the diffusion model during RL iterations. This enhancement arises because opt and filter remove redundant and unstable structures, thereby promoting the diffusion model to discover unexplored material space. All results demonstrate the essential contribution of opt and filter to RL fine-tuning of diffusion models.

\clearpage
\subsubsection{Effect of experience replay} \label{si:sec:ablation:replay}

\begin{figure}[htp]
\centering
\includegraphics[width=1.0\textwidth]{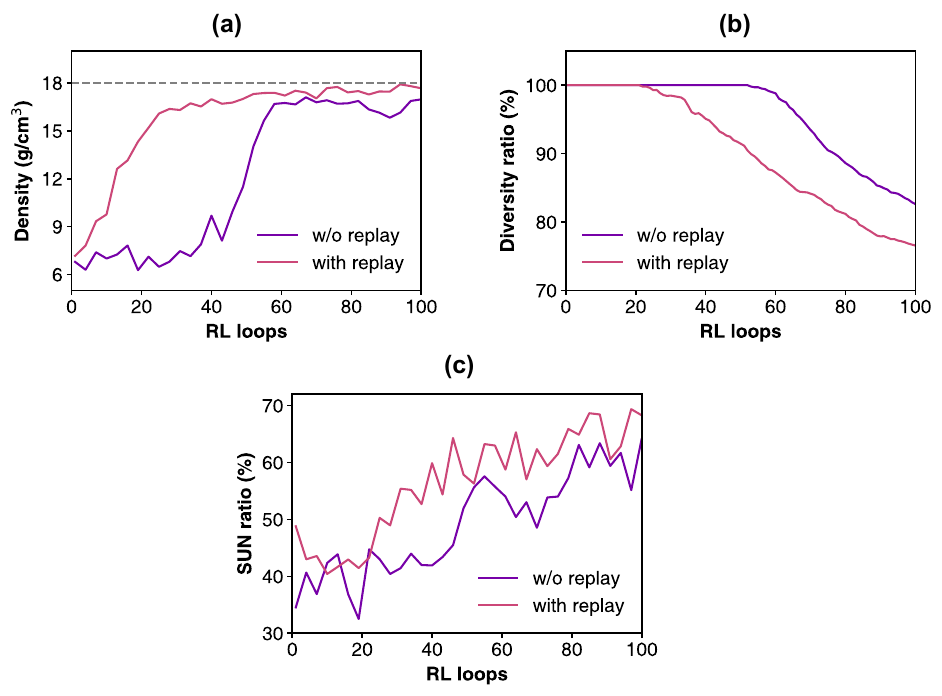}
\caption{
The RL optimization curves (a), composition diversity ratios (b), and SUN ratios (c) of generated structures during the MatInvent process with or without experience replay, for the target density of 18.0 g/cm$^3$.
The optimization curves represents the average values of density of the generated structures in each RL iteration.
}
\label{fig:ablation_replay}
\end{figure}

As shown in Fig. \ref{fig:ablation_replay}a, experience replay clearly enhances RL optimization efficiency, enabling convergence to the target value in fewer iterations. Fig. \ref{fig:ablation_replay}c demonstrates that experience replay exerts negligible influence on the SUN ratio of generated structures. However, experience replay will diminish the compositional diversity of crystal structures generated during RL iterations (Fig. \ref{fig:ablation_replay}b). This motivates the adoption of a diversity filter to counteract the reduction in compositional diversity.

\clearpage
\subsubsection{Effect of diversity filter} \label{si:sec:ablation:df}

\begin{figure}[htp]
\centering
\includegraphics[width=1.0\textwidth]{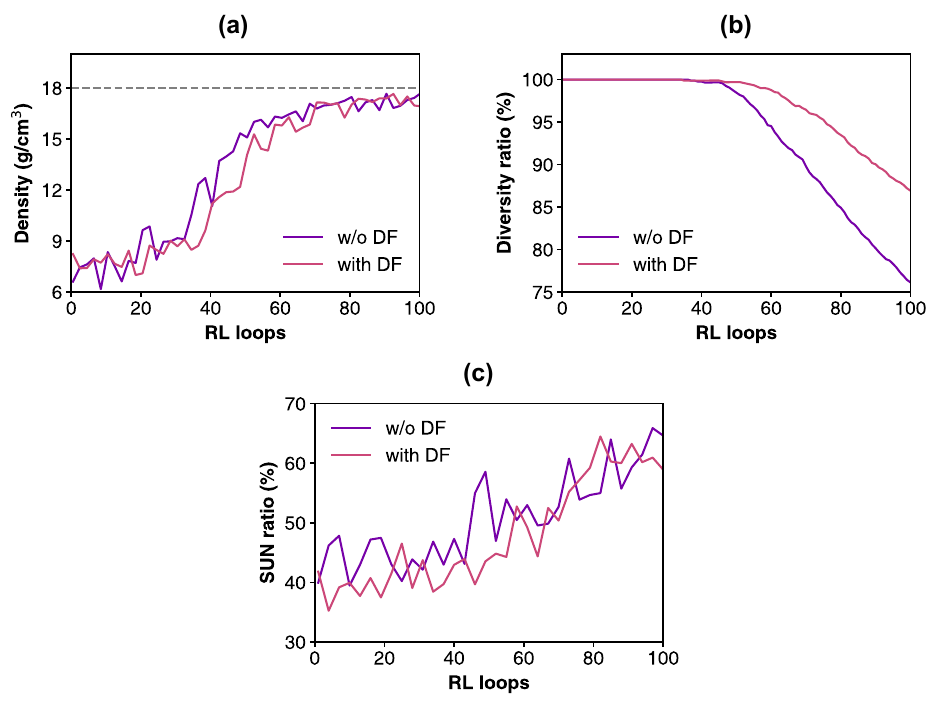}
\caption{
The RL optimization curves (a), composition diversity ratios (b), and SUN ratios (c) of generated structures during the MatInvent process with or without diversity filter (DF), for the target density of 18.0 g/cm$^3$.
The optimization curves represents the average values of density of the generated structures in each RL iteration.
}
\label{fig:ablation_df}
\end{figure}

As shown in Fig. \ref{fig:ablation_df}b, the diversity filter (DF) substantially enhances the compositional diversity of structures generated during RL iterations. This improvement arises because DF penalizes structures with duplicate compositions relative to previously generated samples by assigning lower rewards, thereby incentivizing the diffusion model to explore new chemical compositions and material space. Notably, Fig. \ref{fig:ablation_df}a and c demonstrate that DF exerts negligible influence on both RL optimization efficiency and the SUN ratio of generated crystal structures.

\clearpage
\subsubsection{Effect of the weight of KL regularization}

\begin{figure}[htp]
\centering
\includegraphics[width=1.0\textwidth]{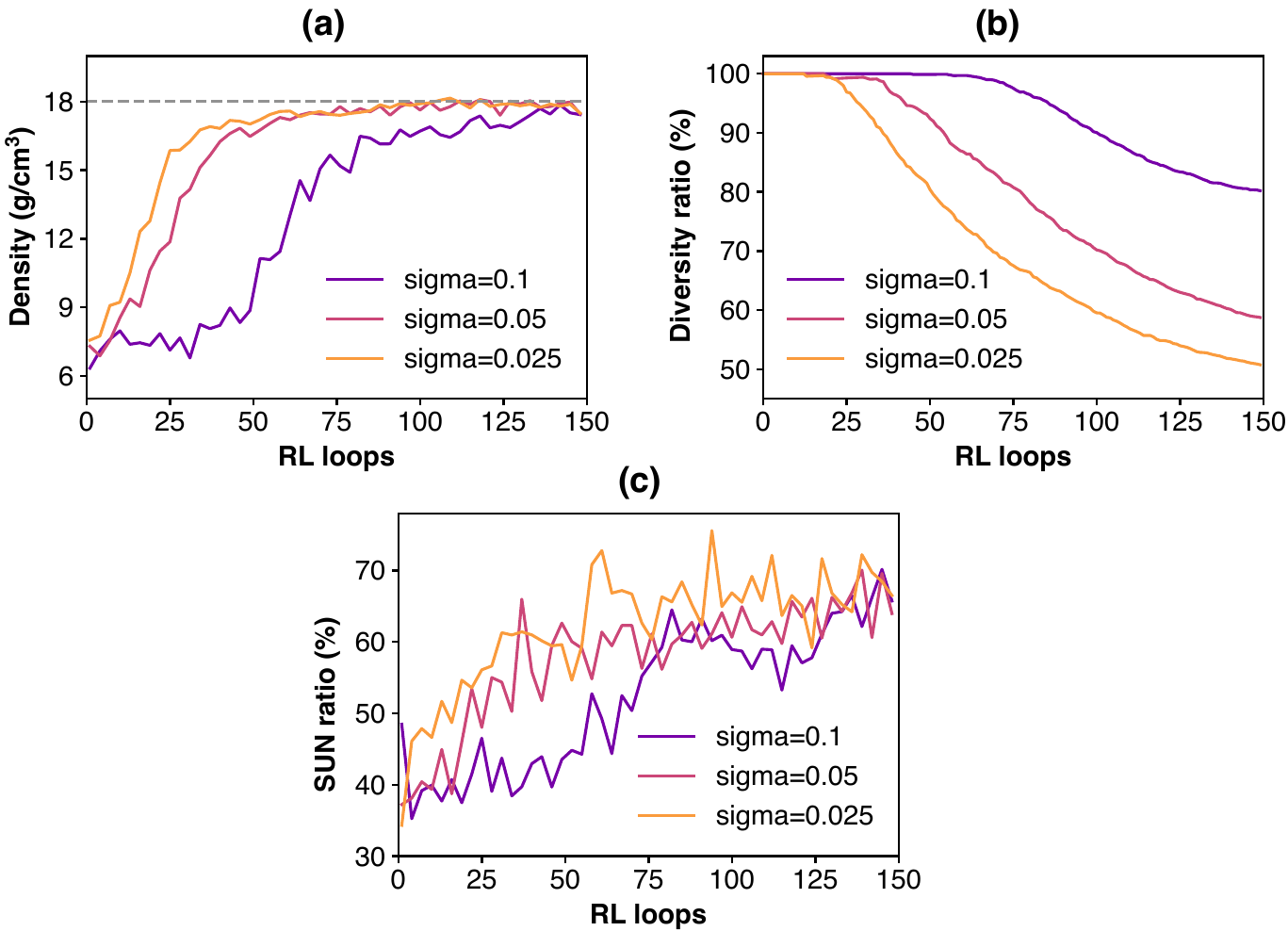}
\caption{
The RL optimization curves (a), composition diversity ratios (b), and SUN ratios (c) of generated structures during the MatInvent process with different weights of KL regularization, for the target density of 18.0 g/cm$^3$.
The optimization curves represents the average values of density of the generated structures in each RL iteration.
}
\label{fig:ablation_sigma}
\end{figure}

The risk of fine-tuning solely based on rewards related to target properties is that the diffusion model may overfit to the rewards and move too far away from the initial state (pre-trained model). To retain the broad material knowledge that the diffusion model has learned from the pre-training dataset for generating reasonable and valid crystal structures, we add the reward-weighted KL between the pre-trained and current fine-tuned models as a regularizer to the RL objective function. $\sigma$ is the weight of KL regularization in the RL objective function. As illustrated in Fig. \ref{fig:ablation_sigma}, a large KL regularization weight enhances compositional diversity but impairs RL optimization efficiency. This behavior mirrors the classic exploration–exploitation trade-off, wherein an appropriate weight achieves an optimal balance between compositional diversity and optimization efficiency. In addition, the absence of KL regularization could lead to failure during RL fine-tuning, as the diffusion model deviates excessively from its pre-trained state and consequently fails to generate chemically reasonable crystal structures.

\clearpage
\section{Material property evaluation} \label{si:sec:property}

\subsection{DFT calculations} \label{sec:dft}

Density functional theory (DFT) calculations were conducted using Vienna Ab initio Simulation Package (VASP) \citeSI{si_vasp1, si_vasp2} with projector augmented wave (PAW) method, accessed via atomate2 \citeSI{si_atomate2} and pymatgen \citeSI{si_pymatgen} software. All computational parameters followed Materials Project \citeSI{si_mp} protocols, including the Perdew–Burke–Ernzerhof (PBE) functional within the generalized gradient approximation (GGA) \citeSI{si_pbe_gga1, si_pbe_gga2}, and Hubbard U corrections \citeSI{si_dftu1}. The workflow for different properties is as follows:

(1) The total energy and energy above hull were calculated by the \texttt{DoubleRelaxMaker} and \texttt{StaticMaker} classes in the atomate2 software \citeSI{si_atomate2} with default settings. Specifically, this workflow includes two back-to-back relaxations and a static calculation.

(2) The band gaps were calculated by the \texttt{RelaxBandStructureMaker} class in the atomate2 software \citeSI{si_atomate2} with default settings. Specifically, this workflow includes two back-to-back relaxations, a static calculation to generate the charge density, a non-self-consistent field calculation on a dense uniform mesh, and a non-self-consistent field calculation on the high-symmetry k-point path to generate the line mode band structure \citeSI{si_dftu1}.

(3) The magnetic densities of generated structures were calculated by the \texttt{DoubleRelaxMaker} and \texttt{StaticMaker} classes in the atomate2 software \citeSI{si_atomate2} with default settings. Specifically, this workflow includes two back-to-back relaxations and a static calculation. The magnetic density is defined as the total magnetization (magnetic moment) of the simulation unit cell divided by the volume of unit cell.

(4) The total dielectric constants were calculated by the \texttt{DoubleRelaxMaker} and \texttt{DielectricMaker} classes in the atomate2 software \citeSI{si_atomate2} with default settings. Specifically, this workflow includes two back-to-back relaxations and a static calculation using density functional perturbation theory to obtain static and high-frequency (ionic) dielectric constants \citeSI{si_mp_dielectric1, si_mp_dielectric2}. Static dielectric constant is electronic contribution to the total dielectric constant. High-frequency (ionic) dielectric constant is ionic contribution to the total dielectric constant. The total dielectric tensor ($3 \times 3$ matrix) can be computed by the ionic ($\epsilon^0$) and electronic ($\epsilon^{\infty}$) contributions: $\epsilon_{i j}=\epsilon_{i j}^0+\epsilon_{i j}^{\infty}$. The total dielectric constant is the average of the diagonal elements of the total dielectric tensor.

For RL experiments using the DFT property evaluation, sample generation and fine-tuning of the diffusion model are performed on the GPU, while all DFT tasks are sent to the CPU cluster and run concurrently to reduce latency. Each DFT task is assigned a maximum computation time limit of 2 hours, the computed property values are sent back to the GPU cluster for RL fine-tuning, while tasks that time out or fail return a None value. Generated structures with a property value of None are deleted, but these DFT computations are still included in the property evaluation cost.

\clearpage
\subsection{ MLIP-based simulations}

\subsubsection{Heat capacity}

The specific heat capacities of the generated materials at 300 K were obtained through geometry optimization and phonon calculations using FairChem software (version 1.10.0) \citeSI{si_oc20}, quacc software \citeSI{si_quacc}, and pre-trained machine learning potentials \texttt{eSEN-30M-OAM} \citeSI{si_omat24, si_escn}. Specifically, the calculation workflow can be divided into the following steps:

(1) runs a relaxation on the unit cell and atoms;

(2) repeats the unit cell a number of times to make it sufficiently large to capture many interesting vibrational models;

(3) generatives a number of finite displacement structures by moving each atom of the unit cell a little bit in each direction;

(4) running single point calculations on each of (3);

(5) gathering all of the calculations and calculating second derivatives (the Hessian matrix);

(6) calculating the eigenvalues/eigenvectors of the Hessian matrix to find the vibrational modes of the material

(7) analyzing the thermodynamic properties of the vibrational modes.

\begin{figure}[htp]
\centering
\includegraphics[width=1.0\textwidth]{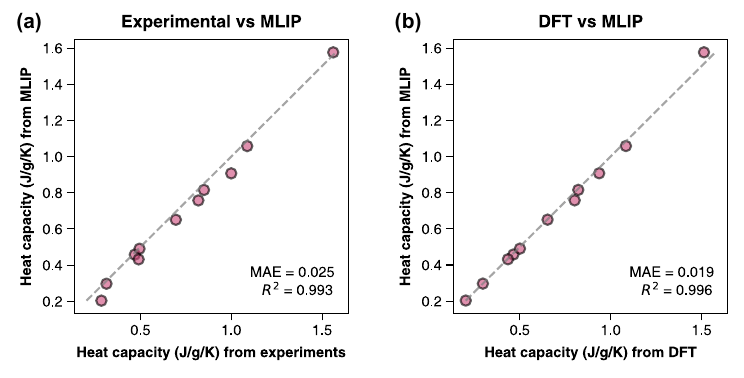}
\caption{
The linear correlation between the specific heat capacity calculated based on MLIP and the experimental (a) or DFT (b) results.
}
\label{fig:heat_mlip}
\end{figure}

As shown in the Fig. \ref{fig:heat_mlip} and Table \ref{table:heat_mlip}, the specific heat capacity calculated based on MLIP has an excellent linear correlation with the experimental results or DFT results \citeSI{si_heat_capacity}. This shows that MLIP simulation can be used as a fast and accurate method to calculate the specific heat capacity in RL.

\begin{table}[th]
  \caption{The linear correlation between the specific heat capacity calculated based on MLIP and the experimental or DFT results.}
  \label{table:heat_mlip}
  \centering
  \begin{threeparttable}
    \begin{tabular}{lccc}
        \hline
        Formula & Experimental & DFT & MLIP \\
        \hline
        KCl & 0.695 & 0.653 & 0.651 \\
        NaCl & 0.850 & 0.822 & 0.815 \\
        ZnS & 0.469 & 0.465 & 0.458 \\
        LiF & 1.562 & 1.513 & 1.58 \\
        ZnO & 0.495 & 0.501 & 0.491 \\
        AlAs & 0.490 & 0.435 & 0.432 \\
        AlN & 0.819 & 0.802 & 0.757 \\
        NaF & 1.088 & 1.084 & 1.058 \\
        PbS & 0.285 & 0.2028 & 0.203 \\
        KI & 0.313 & 0.297 & 0.297 \\
        MgO & 1.0 & 0.937 & 0.907   \\
        \hline
    \end{tabular}
  \end{threeparttable}
\end{table}

\subsubsection{Minimal co-incident area (MCIA)}

Advanced materials synthesis techniques, including Chemical Vapor Deposition (CVD), Molecular Beam Epitaxy (MBE), and sputtering, are widely employed in contemporary materials research. A critical consideration in implementing these techniques is the rational selection of combination of films and substrates. Successful epitaxial growth of heterogeneous interfaces requires multiple factors: the crystallographic properties of both substrate and film materials, preferred cleavage planes, lattice mismatch parameters, and the resulting stress-strain fields at the interface \citeSI{si_mcia1, si_mcia2}.

The Si(100) substrate serves as the industry standard for semiconductor device fabrication due to its superior electronic properties and processing advantages. It is ideal for metal-oxide-semiconductor field-effect transistors (MOSFETs) in modern integrated circuits such as central processing unit (CPU) and graphics processing unit (GPU). Heteroepitaxial growth on Si(100) substrates requires precise control of film thickness and interface quality—critical parameters for device performance.

The minimal co-incident areas (MCIA) between the generated crystal structures (film) and Si(100) substrate were calculated by Zurr \& McGill method \citeSI{si_mcia1, si_mcia2} using MatterSim \citeSI{si_mattersim} MLIP and pymatgen \citeSI{si_pymatgen} software. First, symmetry-preserving geometric relaxations of lattice vectors and atomic coordinates were performed on the conventional cells of generated crystal structures using MatterSim \citeSI{si_mattersim} MLIP. Subsequently, MCIA was calculated from the crystallographic information of the generated structures and Si(100) using the \texttt{SubstrateAnalyzer} class in the pymatgen package \citeSI{si_pymatgen}.

\clearpage
\subsection{ML prediction models}
\subsubsection{Bulk modulus} \label{sec:bulk}

\begin{figure}[htp]
\centering
\includegraphics[width=0.75\textwidth]{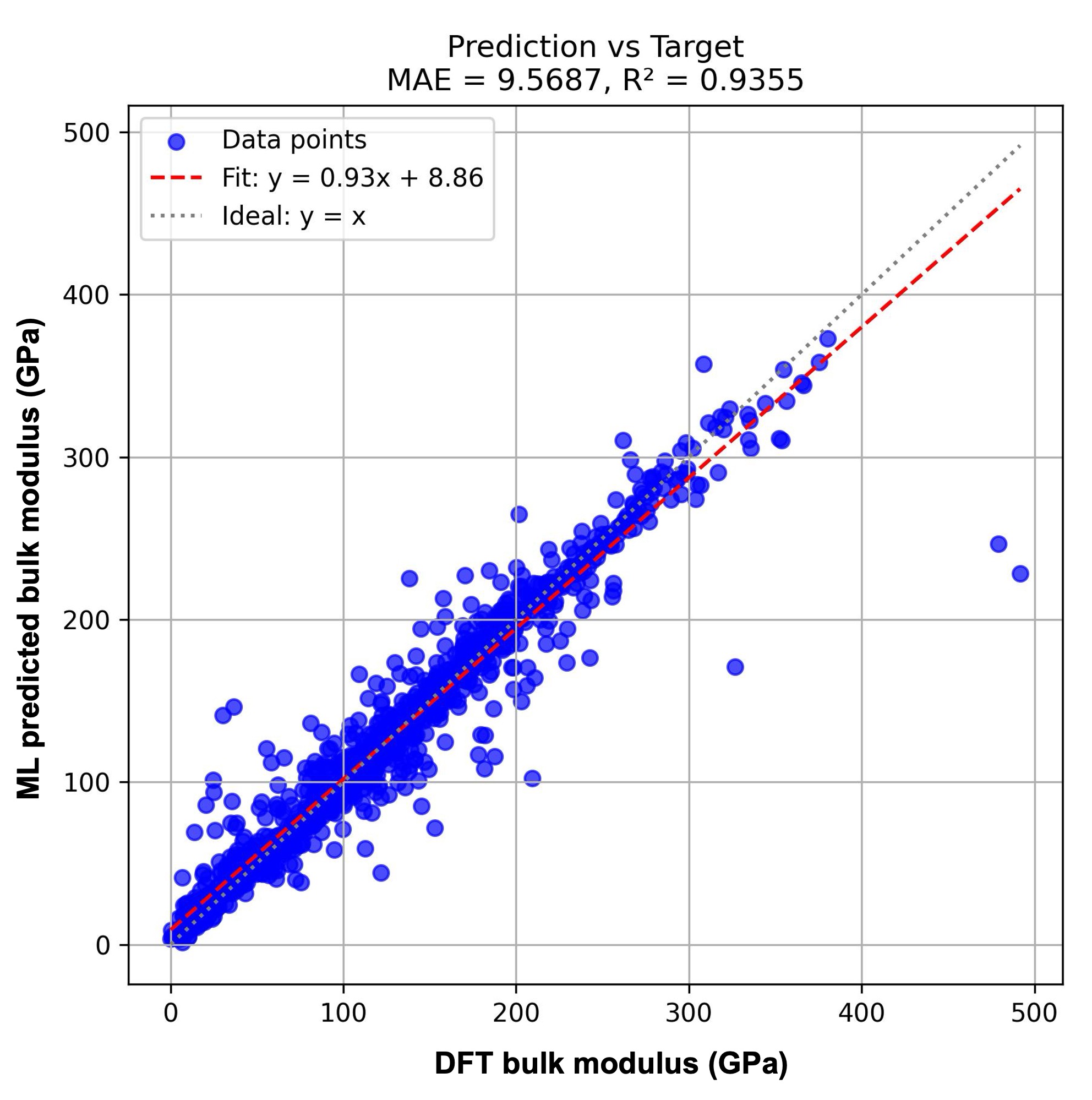}
\caption{
The linear correlation between the bulk modulus predicted by trained ALIGNN model and DFT results on the test set.
}
\label{fig:bulk_alignn}
\end{figure}

The bulk modulus of a substance measures its resistance to a uniform compression. It is defined as the ratio of the infinitesimal pressure increase to the resulting relative decrease of volume. A higher bulk modulus indicates greater resistance to compression, meaning a larger pressure is required to produce a given volume change.

We trained an ALIGNN model \citeSI{si_alignn} to predict the bulk modulus of the generated structures during RL process. First, all materials with 3D structures and DFT Voigt-Reuss-Hill (VRH) average bulk modulus values were extracted from Materials Project database \citeSI{si_mp}, which are 12,845 data points in total. The dataset was randomly split into training, validation, and test sets at a ratio of 8:1:1 for the model training. The model uses a periodic 12-nearest-neighbor graph construction method for training and prediction, with a cutoff radius of 8 $\text{\AA}$ for the construction of neighbor list and bonds (edges). The mean squared error (MSE) loss function was used for model training. The model was trained for 200 epochs using the AdamW optimizer \citeSI{si_adamw} with a normalized weight decay of $10^{-5}$ and a batch size of 32. The learning rate schedule follows the one-cycle policy \citeSI{si_one_policy} with a maximum learning rate of 0.001. The model architecture incorporates initial atom representations (size = 92) derived from the CGCNN framework \citeSI{si_cgcnn}, along with 80 initial bond radial basis function (RBF) features and 40 initial bond angle RBF features. The atom, bond, and angle feature embedding layers generate 64-dimensional inputs for subsequent graph convolution layers. The core network architecture comprises 4 ALIGNN layers and 4 graph convolution (GCN) layers, each with a hidden dimension of 256. The final atom-level representations are aggregated through atom-wise average pooling and subsequently mapped to regression outputs via a single linear transformation layer.

As shown in the Fig. \ref{fig:bulk_alignn}, the trained model achieves a mean absolute error (MAE) of 9.57 GPa and a $R^2$ of 0.935 on the test set, showing a great linear correlation with the DFT-calculated bulk modulus.

\subsubsection{Shear modulus} \label{sec:shear}

\begin{figure}[htp]
\centering
\includegraphics[width=0.75\textwidth]{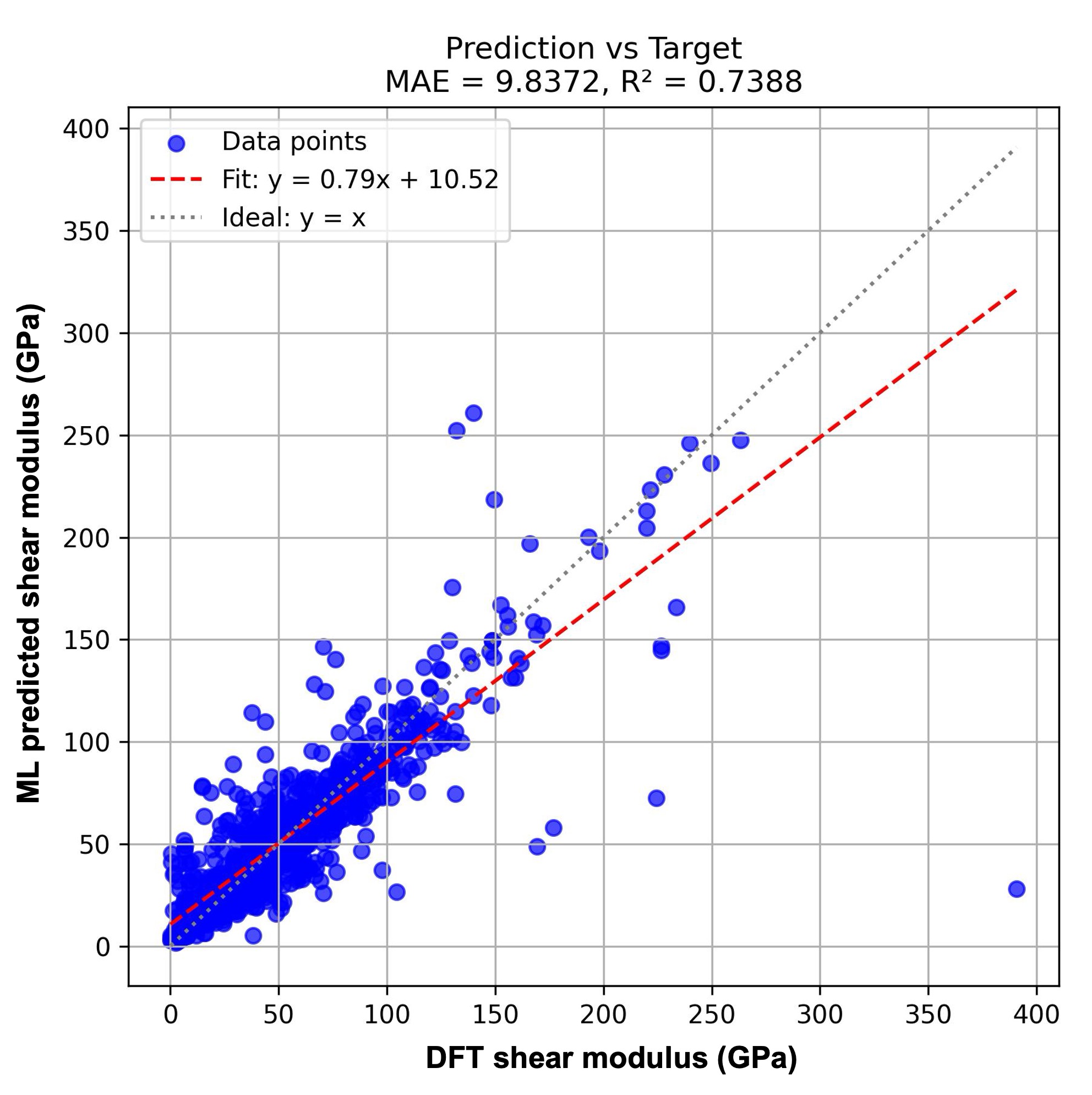}
\caption{
The linear correlation between the shear modulus predicted by trained ALIGNN model and DFT results on the test set.
}
\label{fig:shear_alignn}
\end{figure}

In materials science, shear modulus is a measure of the elastic shear stiffness of a material and is defined as the ratio of shear stress to the shear strain. A higher shear modulus indicates a more rigid material that resists shape changes, while a zero shear modulus signifies a fluid that flows freely. The value is important in fields like structural engineering, material testing, and automotive design, where it helps predict how materials will behave under twisting or shearing forces.

We trained an ALIGNN model \citeSI{si_alignn} to predict the shear modulus of the generated structures during RL process. First, all materials with 3D structures and DFT Voigt-Reuss-Hill (VRH) average shear modulus values were extracted from Materials Project database \citeSI{si_mp}, which are 12,186 data points in total. The dataset was randomly split into training, validation, and test sets at a ratio of 8:1:1 for the model training. The model uses a periodic 12-nearest-neighbor graph construction method for training and prediction, with a cutoff radius of 8 $\text{\AA}$ for the construction of neighbor list and bonds (edges). The mean squared error (MSE) loss function was used for model training. The model was trained for 200 epochs using the AdamW optimizer \citeSI{si_adamw} with a normalized weight decay of $10^{-5}$ and a batch size of 32. The learning rate schedule follows the one-cycle policy \citeSI{si_one_policy} with a maximum learning rate of 0.001. The model architecture incorporates initial atom representations (size = 92) derived from the CGCNN framework \citeSI{si_cgcnn}, along with 80 initial bond radial basis function (RBF) features and 40 initial bond angle RBF features. The atom, bond, and angle feature embedding layers generate 64-dimensional inputs for subsequent graph convolution layers. The core network architecture comprises 4 ALIGNN layers and 4 graph convolution (GCN) layers, each with a hidden dimension of 256. The final atom-level representations are aggregated through atom-wise average pooling and subsequently mapped to regression outputs via a single linear transformation layer.

As shown in the Fig. \ref{fig:shear_alignn}, the trained model exhibits a MAE of 9.84 GPa and a $R^2$ of 0.739 on the test set, showing a good linear correlation with the DFT-calculated shear modulus.

\subsubsection{Young's modulus}

Young's modulus ($E$) quantifies the stiffness of an isotropic elastic material, defined as the ratio of uniaxial stress to strain in the elastic regime:
\begin{equation}
E = \frac{\sigma}{\varepsilon} = \frac{F/A}{\Delta L/L_0}
\end{equation}
where $\sigma$ is stress, $\varepsilon$ is strain, $F$ is force, $A$ is cross-sectional area, $\Delta L$ is elongation, and $L_0$ is original length.

Young's modulus characterizes the resistance to elastic deformation under tensile or compressive loads. High $E$ values indicate high stiffness and small deformations under load (e.g., diamond: $\sim$ 1000 GPa, steel: $\sim$ 200 GPa), suitable for structural applications requiring dimensional stability. Low $E$ values indicate high compliance and large deformations under load (e.g., rubber: about 0.01–0.1 GPa), suitable for flexible or shock-absorbing applications.

Young's modulus can be derived from the bulk modulus ($K$) and shear modulus ($G$):
\begin{equation}
E = \frac{9KG}{3K + G}
\end{equation}
In this work, the Young's modulus was calculated by ML-predicted bulk and shear modulus (Section \ref{sec:bulk} and \ref{sec:shear}).

\subsubsection{Pugh ratio}
The Pugh ratio is a material science criterion calculated by a material's bulk modulus divided by its shear modulus to predict its ductility or brittleness. Materials with a higher Pugh ratio are more likely to be ductile and tough, while materials with a lower ratio tend to be brittle and prone to fracture. This ratio indicates whether a material is more prone to plastic deformation or fracture. In this work, the Pugh ratio was calculated by ML-predicted bulk and shear modulus (Section \ref{sec:bulk} and \ref{sec:shear}).

\subsubsection{Total dielectric constant} \label{si:sec:dielectric}

\begin{figure}[htp]
\centering
\includegraphics[width=0.74\textwidth]{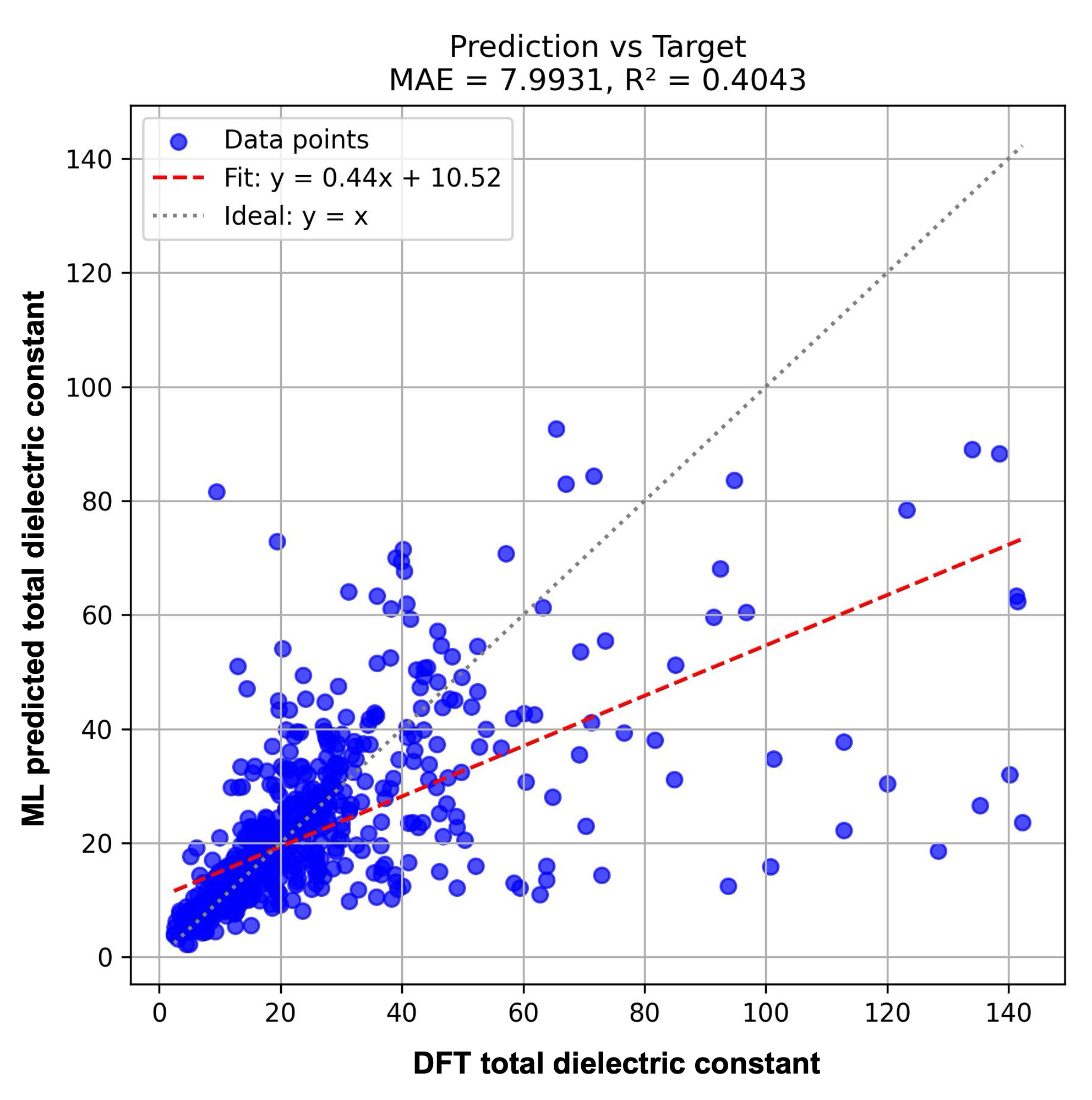}
\caption{
The linear correlation between the total dielectric constant predicted by trained ALIGNN model and DFT results on the test set.
}
\label{fig:dielectric_alignn}
\end{figure}

We trained an ALIGNN model \citeSI{si_alignn} to predict the total dielectric constant of the generated structures during RL process. All crystal structures with DFT-calculated total dielectric constants ranging from 0 to 150 were extracted from Materials Project database \citeSI{si_mp}, which are 7,227 data points in total. The dataset was randomly split into training, validation, and test sets at a ratio of 8:1:1 for the model training. The model uses a periodic 12-nearest-neighbor graph construction method for training and prediction, with a cutoff radius of 8 $\text{\AA}$ for the construction of neighbor list and bonds (edges). The mean squared error (MSE) loss function was used for model training. The model was trained for 200 epochs using the AdamW optimizer \citeSI{si_adamw} with a normalized weight decay of $10^{-5}$ and a batch size of 32. The learning rate schedule follows the one-cycle policy \citeSI{si_one_policy} with a maximum learning rate of 0.001. The model architecture incorporates initial atom representations (size = 92) derived from the CGCNN framework \citeSI{si_cgcnn}, along with 80 initial bond radial basis function (RBF) features and 40 initial bond angle RBF features. The atom, bond, and angle feature embedding layers generate 64-dimensional inputs for subsequent graph convolution layers. The core network architecture comprises 4 ALIGNN layers and 4 graph convolution (GCN) layers, each with a hidden dimension of 256. The final atom-level representations are aggregated through atom-wise average pooling and subsequently mapped to regression outputs via a single linear transformation layer.

As shown in the Fig. \ref{fig:dielectric_alignn}, the trained model exhibits a MAE of 8.0 and a $R^2$ of 0.40 on the test set. The unsatisfactory accuracy may be attributed to the small dataset size and the complex structure-property relationships. Future work should investigate more accurate predictive models \citeSI{si_mao2024dielectric}.

\subsubsection{Formation energy}
Formation energy (unit: eV/atom) is the energy change when one mole of a substance is formed from its constituent elements in their standard states, indicating the material's thermodynamic stability. A negative formation energy signifies that the material is stable and can be formed, while a positive value suggests it is more difficult to form. This parameter is crucial in materials science for designing stable catalysts.

In this work, the formation energy was predicted by ALIGNN model from Ref \citeSI{si_alignn} pretrained on data from Materials Project \citeSI{si_mp}.

\subsection{Synthesizability score}

The synthesizability scores of materials were predicted by the model of Jung et al \citeSI{si_syn_score}. The model was trained by positive-unlabeled learning to predict the likelihood of synthesizing inorganic materials for any given elemental stoichiometries. This model shows a true positive rate of 83.4 \% for the test dataset and an estimated precision of 83.6 \%. The output probability of this model is defined as the synthesizability score, which ranges from 0 to 1. Generally, a score higher than 0.5 indicates that the crystal is likely to be experimentally synthesized.

\subsection{HHI score} \label{sec:hhi}
Herfindahl-Hirschman index (HHI) score based on geological reserves for crystals were calculated by the \texttt{HHIModel} class in the pymatgen package \citeSI{si_pymatgen}. In terms of chemical composition, HHI score based on geological and geopolitical data, provides a quantitative measure of resource economic factors for evaluating the supply and demand risk of materials. It is also a measure of how geographically confined or dispersed the elements comprising a compound are. Using the United States Geological Survey (USGS) commodity statistics, the HHI parameter can be calculated as sum squared of market fraction ($\chi_i$) for a given country, based on their production ($\mathrm{HHI}_{\mathrm{P}}$) or geological reserves ($\mathrm{HHI}_{\mathrm{R}}$) of each element \citeSI{si_hhi, si_hhi2}. Here, for each composition, the weighted average $\mathrm{HHI}_{\mathrm{R}}$ values were calculated using weight fraction of each element in the chemical formula. The U.S. Department of Justice and the Federal Trade Commission define markets as unconcentrated, highly concentrated, or moderately concentrated for a given commodity when HHI scores are below 1500, over 2500, and between these limits, respectively. A lower HHI is desirable, and materials with an HHI score of less than 1500 are considered to have low supply chain risk \citeSI{si_hhi, si_hhi2}.

\subsection{Crustal abundance}
Crustal abundance refers to the concentration of elements of a material in the Earth's crust, typically expressed in parts per million (ppm). Higher crustal abundance indicates greater natural reserves and easier extraction, ensuring long-term supply security. Abundant elements (e.g., Si: $\sim$ 280,000 ppm, Al: $\sim$82,000 ppm) are generally less expensive than rare elements (e.g., In: $\sim$0.05 ppm, rare earth elements: $<$ 100 ppm), directly impacting material production costs. Materials derived from abundant elements are more sustainable for large-scale applications, reducing environmental impact and geopolitical supply risks. Low crustal abundance elements often create supply chain vulnerabilities, particularly for critical technologies (e.g., indium in displays, rare earths in magnets).

The crustal abundance of each element was obtained from the SMACT package, and the crustal abundance of a given material (compound) was calculated using by the weighted average of the crustal abundance (in ppm) based on the mass fraction of each element in the compound:
\begin{equation}
    CA_{\text{compound}} = \sum_{i=1}^{n} CA_i \times w_i
\end{equation}
where $CA_{\text{compound}}$ is the crustal abundance of the compound, $CA_i$ is the crustal abundance of element $i$, $w_i$ is the mass fraction of element $i$ in the compound, and $n$ is the number of elements in the compound.

\subsection{Price}
The cost of a material is one factor that must be considered in its industrial production. To this end, elemental prices obtained from market statistics are used to estimate the raw material costs of a given compound. Low costs and prices are desirable and crucial for sustainable large-scale production.

In this work, \texttt{CostAnalyzer} and \texttt{CostDBElements} classes in pymatgen \citeSI{si_pymatgen} package were used to calculate the price/cost of a given material. The price $P$ (unit: USD/kg) was computed based on the mass fraction and price of each element in a compound:
\begin{equation}
    P = \sum_{i=1}^{n} P_i \times w_i
\end{equation}
where $P_i$ is the price of element $i$, $w_i$ is the mass fraction of element $i$ in the compound, and $n$ is the number of elements in the compound.

\clearpage
\section{Single property optimization}

\subsection{Experimental details} \label{si:sec:spo_exp}

In each RL experiment, the unconditional MatterGen \citeSI{si_mattergen} model pre-trained on Alex-MP-20 dataset \citeSI{si_mattergen} was used as the initial generative model and RL agent. In each RL iteration: 

(1) The diffusion model randomly generates a batch of 64 crystal structures.

(2) The generated structures undergo geometry optimization using \texttt{MatterSim-v1.0.0-5M} MLIP \citeSI{si_mattersim}. Only crystal structures that are thermodynamically Stable ($E_{hull} <$ 0.1 eV/atom), Unique, and Novel (SUN) \citeSI{si_mattergen} are retained. The SUN of each structure was evaluated by Alex-MP reference dataset and MatterGen's code \citeSI{si_mattergen}.

(3) After filtering, 16 samples are randomly selected for property evaluation and assigned corresponding rewards. Due to calculation failure or timeout, the structure with reward of \texttt{None} will be deleted, and this calculation will also be included in the property evaluation budget.

(4) The generated structures with rewards are fed into the diversity filter. The diversity filter imposes a linear penalty on the rewards of structures with non-unique compositions based on the number of previous occurrences. The penalized structures will be removed from the replay buffer (selective memory purge).

(5) The top 50 \% structures ranked by reward and 10 structures randomly sampled with the replay buffer of maximum size 100, are used to fine-tune the diffusion model based on policy optimization with reward-weighted Kullback–Leibler (KL) regularization (weight = 0.025). The learning rate of RL fine-tuning is $10^{-5}$, with a batch size of 16.

(6) To update the replay buffer (maximum size = 100), the top 50 \% structures were added into replay buffer, retaining only the top 100 compositionally unique structures with the highest rewards.

\paragraph{Diversity filter (DF)} In this work. DFs linearly penalize non-unique crystal compositions based on the number of previous occurrences. The reward $r$ is transformed according to the number of previous occurrences (Occ) beyond an allowed tolerance (Tol) until a hard threshold is reached, referred to as the buffer (Buff):
\begin{equation}
r^{\prime}=\left\{\begin{array}{cl}
r \times \frac{\mathrm{Occ}- \text { Tol }}{\text { Buff }- \text { Tol }} & \text { if } \mathrm{Tol}<\mathrm{Occ}<\text { Buff } \\[3pt]
r & \text { if } \quad \text { Occ } \leq \text { Tol } \\[3pt]
0 & \text { if } \quad \text { Occ } \geq \text { Buff }
\end{array}\right.
\end{equation}
where Tol is set to 3 and Buff is set to 6.

\subsection{Reward calculation} \label{si:sec:spo_reward}
In all RL experiments, the values of the target material properties are scaled to between 0 and 1 and used as RL rewards. The property values are derived from DFT calculations, MLIP simulations, or ML prediction models (Section \ref{si:sec:property}). Higher rewards are the optimization goal of the RL process. For tasks involving maximization of target property ($p$) or requiring $p$ to exceed a specified threshold $\tau_1$, the reward $r$ is calculated according to clipped min-max normalization:
\begin{equation}
    r = \text{clip}\left(\frac{p - p_{\min}}{p_{\max} - p_{\min}}, 0, 1\right)
\end{equation}
where $p_{\max}$ and $p_{\min}$ denote the upper and lower bounds of  $p$, respectively, determined by the physically meaningful range of $p$ and the task objectives. The normalized values are clipped to the range [0, 1]. Normally, $p_{\max}$ should be higher than $\tau_1$.

Conversely, for tasks involving minimization of $p$ or requiring $p$ to fall below a given threshold $\tau_2$, the reward $r$ is computed by
\begin{equation}
    r = \text{clip}\left(\frac{p_{\max} - p}{p_{\max} - p_{\min}}, 0, 1\right).
\end{equation}
Normally, $p_{\min}$ should be lower than $\tau_2$.

Moreover, for tasks aimed at achieving a specific target value $\theta$ of $p$, the reward $r$ is given by:
\begin{equation}
    r = \text{clip}\left(\frac{d_{\max} - \left| p - \theta \right|}{d_{\max} - d_{\min}}, 0, 1\right)
\end{equation}
where $d_{\max}$ and $d_{\min}$ denote the upper and lower bounds of the absolute deviation between $p$ and $\theta$, respectively. Normally, $d_{\min}$ is set to 0.

Typically, the MatInvent workflow is configured such that the mean reward at the initial iteration remains below 0.1. The parameters for reward calculation across different tasks are presented as follows:
\begin{itemize}
  \item band gap equal to 3.0 eV: $d_{\min} = 0$ and $d_{\max} = 2$;
  \item magnetic density higher than 0.2 $\text{\AA}^{-3}$: $p_{\min} = 0.0$ and $p_{\max} = 0.25$;
  \item specific heat capacity exceeding 1.5 J/g/K: $p_{\min} = 0.25$ and $p_{\max} = 2.0$;
  \item MCIA below 80 $\text{\AA}^{2}$ on the Si(100) substrate: $p_{\min} = 0.0$ and $p_{\max} = 180$;
  \item bulk modulus of 300 GPa: $d_{\min} = 0$ and $d_{\max} = 250$;
  \item total dielectric constants exceeding 80: $p_{\min} = 35.0$ and $p_{\max} = 120.0$;
  \item synthesizability score higher than 0.9: $p_{\min} = 0.5$ and $p_{\max} = 1.0$;
  \item HHI score below 1250: $p_{\min} = 750$ and $p_{\max} = 3250$;
  \item density of 18.0 g/cm$^3$: $d_{\min} = 0$ and $d_{\max} = 10$.
\end{itemize}

\clearpage
\subsection{More tasks} \label{si:sec:spo_more}
Moreover, MatInvetn was evaluated on six more inverse design tasks with a single target property:
(1) maximizing Young's modulus;
(2) shear modulus of 200 GPa;
(3) maximizing Pugh ratio;
(4) minimizing formation energy;
(5) maximizing crustal abundances of elements in materials;
and (6) minimizing the price of materials.
The description and evaluation methods of target material properties were listed in Section \ref{si:sec:property}.

\begin{figure}[htp]
\centering
\includegraphics[width=1.0\textwidth]{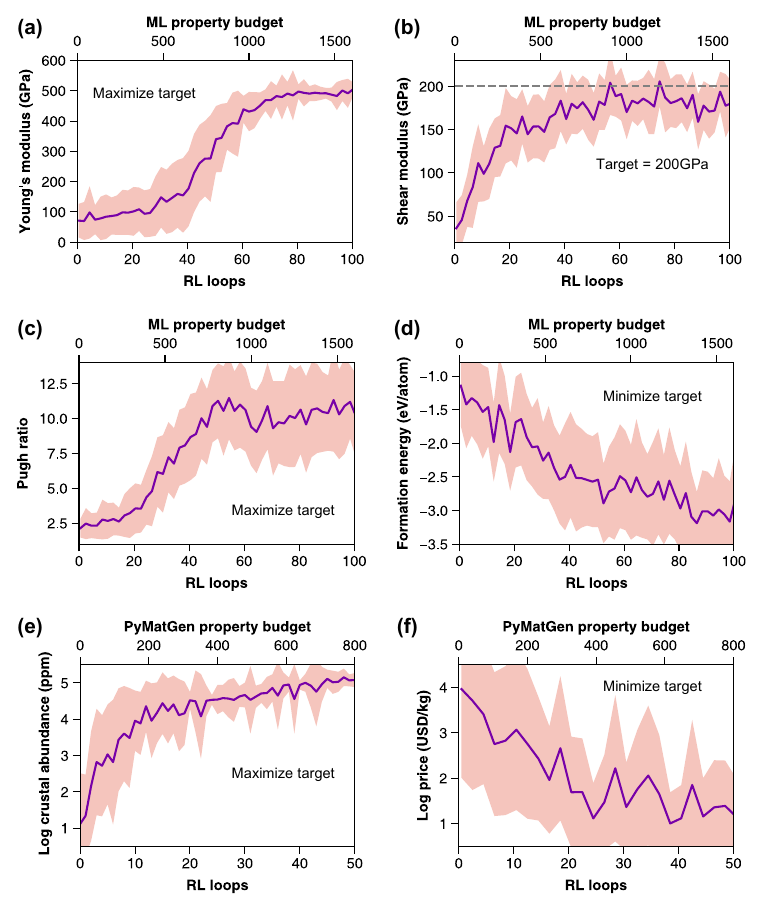}
\caption{
The optimization curves of MatInvent workflow on different inverse design tasks with a single target property:
(a) maximize Young's modulus;
(b) shear modulus of 200 GPa;
(c) maximize Pugh ratio;
(d) minimize formation energy;
(e) maximize crustal abundances of elements in materials;
and (f) minimize the price of materials.
The curves represents the average values of the target properties of the generated structures in each RL iteration, while the shading depicts standard deviation.
}
\label{fig:other_spo_tasks}
\end{figure}

Across all tasks (Fig. \ref{fig:other_spo_tasks}), as RL iterations progressed, the average property values of the generated materials continued to move toward the target. Notably, after 50 iterations (approximately 800 property evaluation costs), the average property values were close to convergence. All results demonstrate that our MatInvent is a general and highly efficient RL framework tailored for diffusion models in single property optimization tasks.

\clearpage
\subsection{Property distributions} \label{si:sec:spo_dis}
\begin{figure}[htp]
\centering
\includegraphics[width=1.0\textwidth]{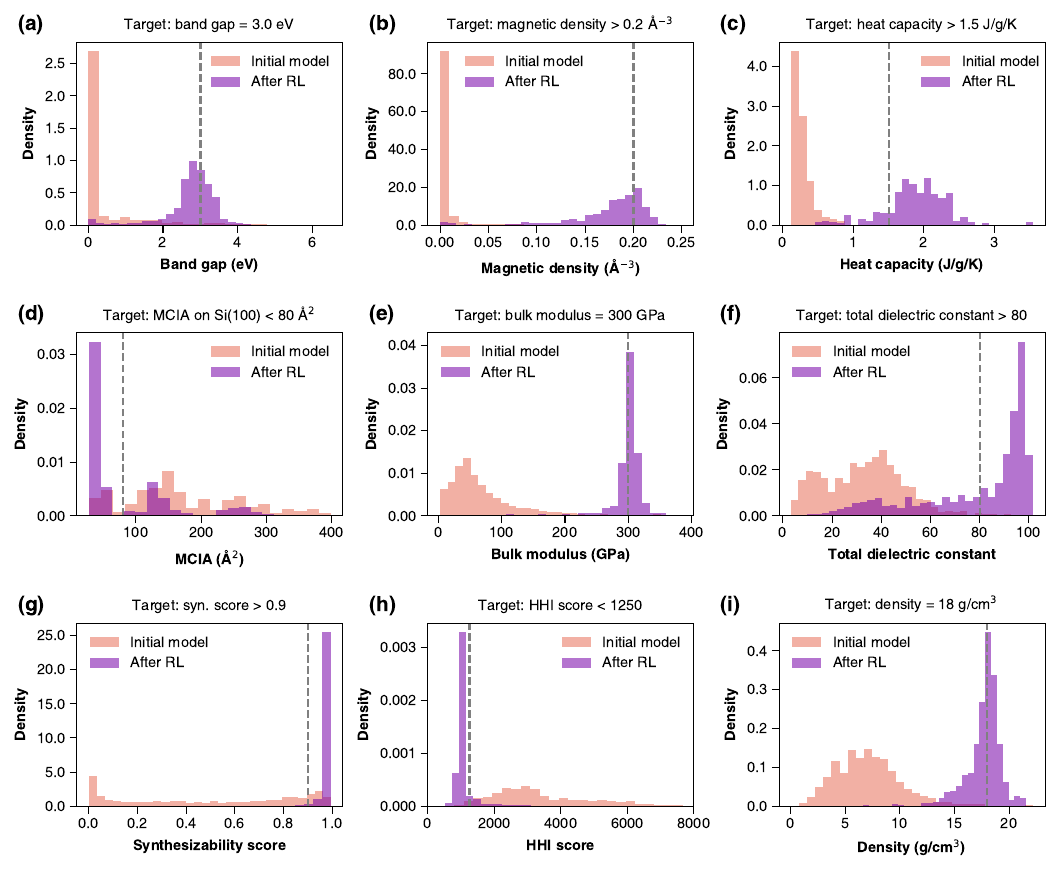}
\caption{Probability density distributions of property values of SUN structures generated by pretrained and RL-finetuned diffusion models for inverse design targets of
(a) band gap equal to 3.0 eV;
(b) magnetic density higher than 0.2 $\text{\AA}^{-3}$;
(c) specific heat capacity exceeding 1.5 J/g/K;
(d) MCIA below 80 $\text{\AA}^{2}$ on the Si(100) substrate;
(e) bulk modulus of 300 GPa;
(f) total dielectric constants exceeding 80;
(g) synthesizability score higher than 0.9;
(h) HHI score below 1250;
and (i) density of 18.0 g/cm$^3$.
}
\label{fig:spo_dis}
\end{figure}

 In all RL experiments, the unconditional MatterGen \citeSI{si_mattergen} model pre-trained on Alex-MP-20 dataset was utilized as the initial model of RL process. In each task in Figure 3, the initial model and the RL-finetuned model after 100 iterations generated 1024 structures respectively. For Fig. \ref{fig:spo_dis} a and b, all generated structures were relaxed using the MatterSim \citeSI{si_mattersim} MLIP and filtered by stability, uniqueness, and novelty. Subsequently, for each model, 150 structures were randomly selected and evaluated by DFT calculations (Section \ref{sec:dft}). For Fig. \ref{fig:spo_dis} c-i, all generated structures were relaxed using the MatterSim \citeSI{si_mattersim} MLIP, and only SUN structures were evaluated for target properties using MLIP simulation, ML prediction and pymatgen \citeSI{si_pymatgen} (Section \ref{si:sec:property}).

\clearpage
\subsection{SUN ratio after RL finetuning} \label{si:sec:spo_sun}
\begin{figure}[htp]
\centering
\includegraphics[width=0.85\textwidth]{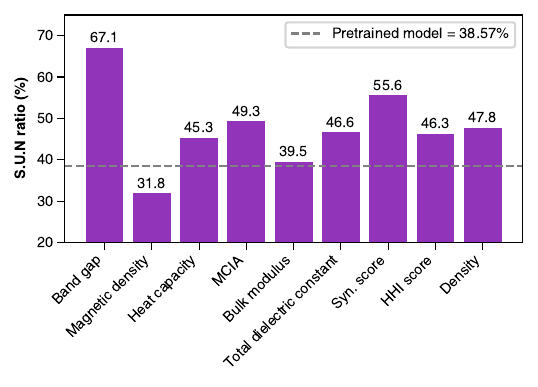}
\caption{
SUN ratios of 1024 generated structures by RL-finetuned diffusion models for different inverse design targets:
band gap equal to 3.0 eV;
magnetic density higher than 0.2 $\text{\AA}^{-3}$;
specific heat capacity exceeding 1.5 J/g/K;
minimal co-incident area (MCIA) below 80 $\text{\AA}^{2}$ on the Si(100) substrate;
bulk modulus of 300 GPa;
total dielectric constants exceeding 80;
synthesizability score higher than 0.9;
Herfindahl–Hirschman index (HHI) score below 1250;
and density of 18.0 g/cm$^3$.
}
\label{fig:sun_rl}
\end{figure}

We found that the SUN ratio of MatterGen \citeSI{si_mattergen} after conditional generation decreased compared to the unconditional pre-trained model. For example, the SUN ratio of conditional generation at a magnetic density of 0.2 $\text{\AA}^{-3}$ was 13.1 \%, a decrease of approximately 25 \%. This may be a shortcoming of conditional generation.
As illustrated in Fig. \ref{fig:sun_rl}, most RL fine-tuned models exhibited higher SUN ratios ($>$ 45 \%) relative to the initial pretrained model (38.7 \%), which can be attributed to MLIP-based structure optimization and SUN filtering prior to property evaluation.

\clearpage
\subsection{Comparison between MatterGen conditional generation and MatInvent} \label{si:sec:spo_mattergen}
For a fair comparison, our RL experiments utilized the same unconditional MatterGen model pre-trained on Alex-MP-20 dataset as the initial model \citeSI{si_mattergen}.
For the task targeting materials with magnetic densities exceeding 0.2 $\text{\AA}^{-3}$, MatterGen generated 4096 samples with their fine-tuned model by conditioning on a magnetic density value of 0.2 $\text{\AA}^{-3}$ \citeSI{si_mattergen}. Similarly, the RL-finetuned diffusion model after 100 iterations using the MatInvent workflow also generated 4096 structures. All structures were relaxed using the MatterSim \citeSI{si_mattersim} MLIP and filtered by stability, uniqueness, and novelty. Subsequently, for each model, 250 structures were randomly selected and subjected to DFT evaluation.

For the task targeting materials with bandgaps of 3.0 eV, MatterGen generated 1024 structures with their fine-tuned model by conditioning on a value of 3.0 eV for band gap \citeSI{si_mattergen}. Similarly, the RL-finetuned diffusion model after 100 iterations also generated 1024 structures. All structures were relaxed using the MatterSim \citeSI{si_mattersim} MLIP and filtered by stability, uniqueness, and novelty. Subsequently, for each model, 250 structures were randomly selected and subjected to DFT evaluation.

\clearpage
\section{Multiple property optimization}

\subsection{Experimental details}
For the task targeting materials with magnetic densities exceeding 0.2 $\text{\AA}^{-3}$ and HHI score below 1500, MatInvent followed the same hyperparameters and settings as in Section \ref{si:sec:spo_exp}, where the property evaluation size for each RL iteration is 16. The maximum number
of RL iterations is set to 120. The DFT method (Section \ref{sec:dft}) was employed to determine the magnetic density of the generated structures, and pymatgen \citeSI{si_pymatgen} package was utilized to compute their HHI scores (Section \ref{sec:hhi}).

For the task designing novel high-$\kappa$ dielectrics, MatInvent uses a similar setup as in Section \ref{si:sec:spo_exp}, but with a property evaluation size of 32 and a sampling size of 128 per RL iteration. The maximum number of RL iterations is set to 240. Due to computational expense of DFT property evaluation, ML models were employed to predict $E_g$, $\varepsilon_{\text {total }}$, and corresponding FoM of the generated structures during the RL process.

\clearpage
\subsection{Reward calculation} \label{si:sec:mpo_reward}
In the multiple property optimization, the numerical values of material properties are first scaled to the range of 0 to 1 through the clipped Min-Max normalization (Section \ref{si:sec:spo_reward}), and subsequently, the scaled values of different properties are combined to calculate the final reward. Additionally, standardization methods could be investigated in future studies as an alternative approach for scaling the numerical values of material properties.

(1) In the task targeting materials with magnetic densities exceeding 0.2 $\text{\AA}^{-3}$ and HHI score below 1500, the scaled value ($s_m$) of magnetic density ($p_m$) was calculated by
\begin{equation}
    s_m = \text{clip}\left(\frac{p_m}{0.25}, 0, 1\right).
\end{equation}
And the scaled value ($s_h$) of HHI score ($p_h$) was computed by
\begin{equation}
    s_h = \text{clip}\left(\frac{3250 - p_h}{3250 - 750}, 0, 1\right).
\end{equation}
The final reward $r$ was calculated by
\begin{equation}
    r = \min(s_m, s_h).
\end{equation}

(2) For the task designing novel high-$\kappa$ dielectrics, the scaled value ($s_g$) of band gap ($E_g$) was calculated by
\begin{equation}
    s_g = \text{clip}\left(\frac{E_g - 0.5}{3.5 - 0.5}, 0, 1\right).
\end{equation}
And the scaled value ($s_t$) of total dielectric constant ($\varepsilon_{\text{total}}$) was calculated by
\begin{equation}
    s_t = \text{clip}\left(\frac{\varepsilon_{\text{total}} - 25}{50 - 25}, 0, 1\right).
\end{equation}
And the scaled value ($s_f$) of figure of merit ($\text{FoM}=E_g \times \varepsilon_{\text{total}}$) was calculated by
\begin{equation}
    s_f = \text{clip}\left(\frac{\text{FoM} - 10}{250 - 10}, 0, 1\right).
\end{equation}
The final reward $r$ was calculated by
\begin{equation}
    r = \alpha s_g + \beta s_t + (1 - \alpha - \beta) s_f
\end{equation}
where $\alpha = \beta = 0.1$ normally. In addition, $\alpha = \beta = 0$ is also an effective parameter for Pareto optimization.

The trained ALIGNN model \citeSI{si_alignn} in Section \ref{si:sec:dielectric} was used to predict the total dielectric constants of the generated structures during RL process. In addition, we trained an ALIGNN model \citeSI{si_alignn} to predict the band gap of the generated structures during RL process. First, all crystal structures with DFT band gap values were extracted from Materials Project database \citeSI{si_mp}, which are 154,839 data points in total. The dataset was randomly split into training, validation, and test sets at a ratio of 8:1:1 for the model training. The model uses a periodic 12-nearest-neighbor graph construction method for training and prediction, with a cutoff radius of 8 $\text{\AA}$ for the construction of neighbor list and bonds (edges). The mean squared error (MSE) loss function was used for model training. The model was trained for 150 epochs using the AdamW optimizer \citeSI{si_adamw} with a normalized weight decay of $10^{-5}$ and a batch size of 64. The learning rate schedule follows the one-cycle policy \citeSI{si_one_policy} with a maximum learning rate of 0.001. The model architecture incorporates initial atom representations (size = 92) derived from the CGCNN framework \citeSI{si_cgcnn}, along with 80 initial bond radial basis function (RBF) features and 40 initial bond angle RBF features. The atom, bond, and angle feature embedding layers generate 64-dimensional inputs for subsequent graph convolution layers. The core network architecture comprises 4 ALIGNN layers and 4 graph convolution (GCN) layers, each with a hidden dimension of 256. The final atom-level representations are aggregated through atom-wise average pooling and subsequently mapped to regression outputs via a single linear transformation layer.

\begin{figure}[htp]
\centering
\includegraphics[width=0.73\textwidth]{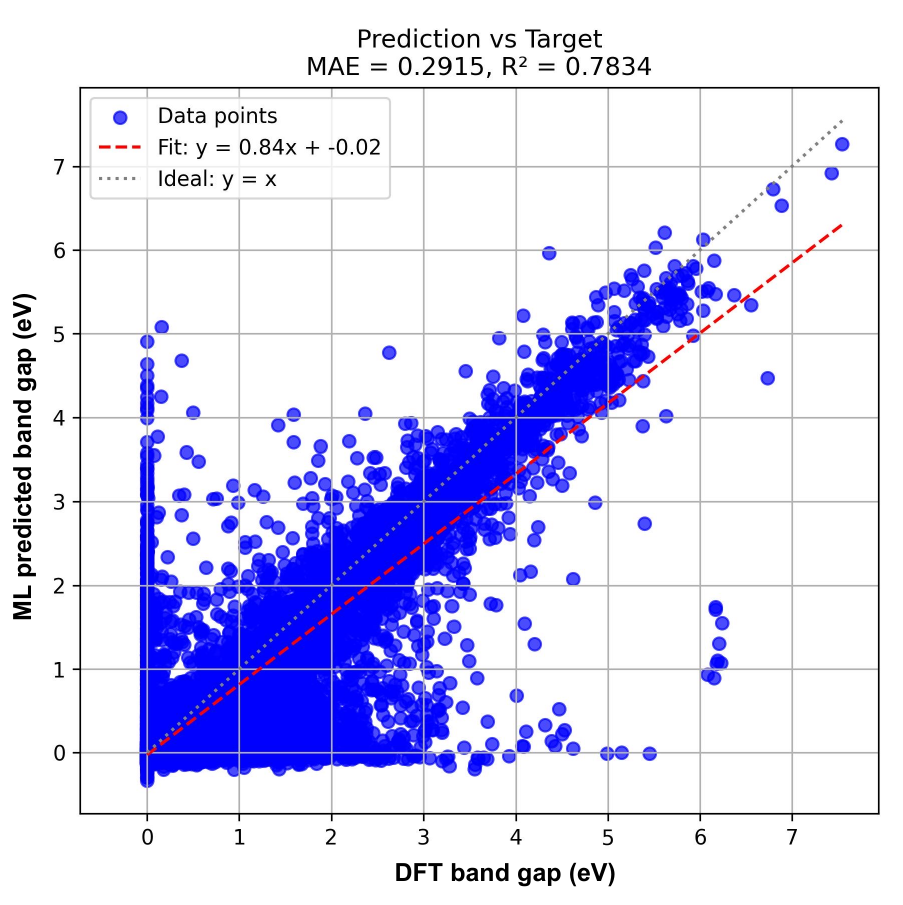}
\caption{
The linear correlation between the band gaps predicted by trained ALIGNN model and DFT results on the test set.
}
\label{fig:gap_alignn}
\end{figure}

As shown in the Fig. \ref{fig:gap_alignn}, the trained model achieves a mean absolute error (MAE) of 0.29 eV and a $R^2$ of 0.78 on the test set, showing a good linear correlation with the DFT-calculated band gap.

\clearpage
\subsection{Comparison between MatterGen conditional generation and MatInvent}  \label{si:sec:mpo_mattergen}
For a fair comparison, our RL experiments utilized the same unconditional MatterGen model pre-trained on Alex-MP-20 dataset as the initial model \citeSI{si_mattergen}.
For the task targeting materials with magnetic densities exceeding 0.2 $\text{\AA}^{-3}$ and HHI score below 1500, MatterGen generated 10,240 structures with their fine-tuned model by jointly conditioning on a magnetic density value of 0.2 $\text{\AA}^{-3}$ and an HHI score of 1500 \citeSI{si_mattergen}. Similarly, the RL-finetuned diffusion model after 120 iterations using the MatInvent workflow also generated 10,240 structures. All structures were relaxed using the MatterSim \citeSI{si_mattersim} MLIP and filtered by stability, uniqueness, and novelty. Subsequently, the HHI scores of all structures were calculated using the pymatgen package \citeSI{si_pymatgen}, and structures with HHI scores higher than 1500 were removed. Finally, for each model, 200 structures were randomly selected and subjected to DFT evaluation.

\clearpage
\bibliographystyleSI{unsrt}
\bibliographySI{si_ref}

\end{document}